\documentclass
[preprint,prb,a4paper,amsfonts,amssymb,floats,titlepage,fleqn,footinbib,lengthcheck,twocolumn,tightenlines,balancelastpage,10pt,showpacs,runinaddress]{revtex4-1}%
\usepackage{natbib}
\usepackage{amssymb}
\usepackage{amsfonts}
\usepackage{amsmath}
\usepackage{graphicx}
\usepackage[usenames,dvipsnames]{color}
\usepackage{longtable}
\usepackage{revsymb}
\usepackage{units}
\usepackage{nicefrac}
\usepackage{textcomp}
\usepackage{epsfig}
\usepackage{footmisc}
\usepackage{soul}%
\ifx\pdfoutput\relax\let\pdfoutput=\undefined\fi
\newcount\msipdfoutput
\ifx\pdfoutput\undefined\else
\ifcase\pdfoutput\else
\ifx\paperwidth\undefined\else
\ifdim\paperheight=0pt\relax\else\pdfpageheight\paperheight\fi
\ifdim\paperwidth=0pt\relax\else\pdfpagewidth\paperwidth\fi
\fi\fi\fi
\begin{document}
\title{Enhanced superconductivity, Kondo behavior and negative-curvature resistivity
of oxygen-irradiated thin films of aluminium}
\author{E. H. C. P. Sinnecker}
\affiliation{Instituto de F\'{\i}sica, Universidade Federal do Rio de Janeiro, Caixa Postal
68528, Rio de Janeiro, Rio de Janeiro 21941-972, Brazil}
\author{M. M. Sant'Anna}
\affiliation{Instituto de F\'{\i}sica, Universidade Federal do Rio de Janeiro, Caixa Postal
68528, Rio de Janeiro, Rio de Janeiro 21941-972, Brazil}
\author{M. ElMassalami}
\affiliation{Instituto de F\'{\i}sica, Universidade Federal do Rio de Janeiro, Caixa Postal
68528, Rio de Janeiro, Rio de Janeiro 21941-972, Brazil}

\begin{abstract}
We followed the evolution of the normal and superconducting properties of Al
thin films after each session of various successive oxygen irradiations at
ambient temperature. Such irradiated films,\ similar to the granular ones,
exhibit enhanced superconductivity, Kondo behavior and negative-curvature
resistivity. Two distinct roles of oxygen are identified: as a damage-causing
projectile and as an implanted oxidizing agent. The former gives rise to the
processes involved in the conventional recovery stages. The latter, considered
within the context of the Cabrera-Mott model, gives rise to a multistep
process which involves charges transfer and creation of stabilized vacancies
and charged defects. Based on the outcome of this multistep process, we
consider (i) the negative curvature resistivity as a manifestation of a
thermally-assisted liberation of trapped electric charges, (ii) the Kondo
contribution as a spin-flip {scattering from }paramagnetic, color-center-type
defects, and (iii) the enhancement of $T_{\text{c}}$ as being due to a lattice
softening facilitated by the stabilized defects and vacancies. The similarity
in the phase diagrams of granular and irradiated films as well as the aging
effects are discussed along the same line of reasoning.

\end{abstract}

\maketitle

\section{Introduction}

Generally, controlled incorporation of a chemically-active element into a
metallic superconducting thin film induces a drastic modification in its
normal and superconducting phase
diagram.\cite{Abeles75-Granular-Films-Review,Deutscher08-Nano-Granular-SUC-Review}
The modification in the superconductivity can be illustrated by the increase
of the transition points, $T_{c}$, of oxygen-incorporated Al films by up to a
factor of
three.{\cite{Bachar15-Mott-granular-Al,*Bachar13-Kondo-granular-Al,*Pracht15-SUC-dome-Granular-Al,Bachar14-PhD-Thesis,Abeles67-Hc-granular-SUCs,Lamoise75-Irradiation-Al-SUC,*Lamoise76-Enhanced-SUC-Al-Alloys,*Lamoise75-Irradiation-Al-Resistivity-annealing,*Meunier77-Irradiation-Al-Percolation-SUC}%
\thinspace\ }The modifications in the normal state, on the other hand, are not
less spectacular: (i) Although both Al and O are nonmagnetic,
oxygen-incorporated Al film exhibits a ($\sim$10 K) Kondo behavior that
competes with the
superconducting\ state\cite{Bachar15-Mott-granular-Al,*Bachar13-Kondo-granular-Al,*Pracht15-SUC-dome-Granular-Al,Bachar14-PhD-Thesis}
and (ii) its resistivity exhibits a negative curvature (NCR) at $\sim$300 K
with a deviation downwards away from the \textit{linear-in-T}
behavior.\cite{Bachar15-Mott-granular-Al,*Bachar13-Kondo-granular-Al,*Pracht15-SUC-dome-Granular-Al,Bachar14-PhD-Thesis}%

It is remarkable that (i) these induced features are evident in both the
granular
films\cite{Abeles75-Granular-Films-Review,Deutscher08-Nano-Granular-SUC-Review,Bachar15-Mott-granular-Al,*Bachar13-Kondo-granular-Al,Bachar14-PhD-Thesis}
(oxygen is incorporated during the codeposition process) and irradiated
films\cite{Lamoise75-Irradiation-Al-SUC,*Lamoise76-Enhanced-SUC-Al-Alloys,*Lamoise75-Irradiation-Al-Resistivity-annealing,*Meunier77-Irradiation-Al-Percolation-SUC}
(oxygen is implanted posterior to film synthesis), (ii) the $T_{c}$
enhancement follows a dome-like
behavior{\cite{Bachar15-Mott-granular-Al,*Bachar13-Kondo-granular-Al,*Pracht15-SUC-dome-Granular-Al,Bachar14-PhD-Thesis,Deutscher73-Granular-SUC-Films,*Deutscher73-Granular-SUC-zeroDimension}%
} reminiscent of the case of HTc cuprates, and (iii) the NCR feature is unique
and has no resemblance to the recovery stages usually observed in
\textit{pure} Al
films.\cite{Isebeck66-Reconvery-in-Al-NeutronIrradiation,*Sosin63-Aluminum-Recovery-AfterIrradiation,*DeSorbo59-Kinetics-Vacancy-Al,*Khellaf02-Quench-Aluminum}%

Historically, the induced features of granular films had been treated
separately from those appearing in irradiated ones. Similarly, each feature
had been treated as if independent of the others. Moreover, none of them had
been correlated with the kinetics of defects (interstitials, vacancies, etc.
see Fig.\ref{Fig1-Al-Kinetics-Cabrera-Mott}) even when these are introduced by
such a damage-causing irradiation process. As a result of these historical
approaches, much of the essential features (and their driving mechanisms) of
the normal and superconducting phase diagrams of the irradiated and granular
films are not well clarified.

In this work we\ systematically studied the evolution of the normal and
superconducting properties of Al thin film when oxygen is progressively
incorporated via irradiation at ambient temperature. A phase diagram is
constructed from the events manifested in the resistivity curves. We analyzed
such a phase diagram and identified the region of operation for each of NCR,
the Kondo behavior and the enhancement of $T_{c}$. We discuss the mechanisms
involved in each effect.

Our experimental methods and materials are presented in Sec. II, the results
and analysis are in Sec. III, while the discussion and a summary are given in
Sec. IV.

\section{Experimental}

Thin films of 400\thinspace$\mu$m$\times$10\thinspace$\mu$m$\times
$90\thinspace$n$m were prepared at room temperature by sputtering Al\ on a
lithographed Si/SiO$_{2}$ substrate with Ti/Au contact pads. Electron
microscopy images indicate a grain size of $\sim$60\thinspace$n$m. Such thin
films were irradiated\cite{Mello16-Accelerator-SIMS} with O$^{-}$ ions at room
temperature; seven consecutive implantation sessions were carried out (see
Table \ref{TabI-Fluence-SRIM}). We observed no significant dependence of the
studied properties on the implantation depth profile when using energies of
10, 23, or 30 keV.

The influence of each irradiation was followed by DC/AC four-point resistivity
measurements both \textit{in situ,} during irradiations, as well as \textit{ex
situ. }The latter ones were measured as a function of time, temperature,
magnetic field and fluence after each $n\mathrm{th}$ irradiation session:
$\rho(t,T,H$, $n\mathrm{th})$. Routinely, $\rho(T,H$,$n\mathrm{th})$ was
measured, directly after $n\mathrm{th}$ irradiation, during the cooling down
to $\sim$1.7K and, afterwards, the warming up to $\sim$320K: Except for aging
effects, the measurements were reproducible and the obtained curves compare
favorably with the ones reported for
granular\cite{Abeles75-Granular-Films-Review,Deutscher08-Nano-Granular-SUC-Review,Bachar15-Mott-granular-Al,*Bachar13-Kondo-granular-Al,Bachar14-PhD-Thesis}
and irradiated
films.\cite{Lamoise75-Irradiation-Al-SUC,*Lamoise76-Enhanced-SUC-Al-Alloys,Lamoise75-Irradiation-Al-Resistivity-annealing,Meunier77-Irradiation-Al-Percolation-SUC}
Hall measurements on representative samples confirmed the earlier
findings\cite{Bandyopadhyay82-HallEffect-granular-Al} that the major charge
carriers are electrons and their density decreases with irradiation.

Based on analysis of thermal evolution of each $\rho(t,T,H$, $x)$ curve, we
identified all transition and crossover events. A plot of these temperature
points versus $x$ gives the $T-x$ phase diagram of the O-irradiated thin films
($x$ is the tunable parameter). We include in this very same phase diagram
\textit{all} the previously reported $T(x)$ of
granular\cite{Abeles75-Granular-Films-Review,Deutscher08-Nano-Granular-SUC-Review,Bachar15-Mott-granular-Al,*Bachar13-Kondo-granular-Al,Bachar14-PhD-Thesis}
and irradiated thin
films.\cite{Lamoise75-Irradiation-Al-SUC,*Lamoise76-Enhanced-SUC-Al-Alloys,Lamoise75-Irradiation-Al-Resistivity-annealing,Meunier77-Irradiation-Al-Percolation-SUC}
It is worth mentioning that many of the reported $T_{c}(x)$ values were given
as a function of $\rho_{\text{300K}}$ (see e.g.
Refs.{\onlinecite{Bachar15-Mott-granular-Al,*Bachar13-Kondo-granular-Al,*Pracht15-SUC-dome-Granular-Al,Bachar14-PhD-Thesis}).
}Here, we maintain the same convention which, due to relaxation effects,
corresponds to our extrapolated $\rho_{\text{300K}}^{ext}$: At any rate this
is essentially a parameter which can be substituted, with no loss of
generality, by $x$ or $\rho_{\text{o}}$ (the latter tracks the combined
influence of the parameters appearing in $\rho_{\text{o}}=m/n\tau e^{2}$; all
terms have their usual meanings).%

\begin{table*}[tbp] \centering
\caption{Beam energy, partial and accumulated fluence, and peak center of depth profile for each of
the consecutive implantation sessions used in this work. The specific range of energies were chosen to probe
any dependence on implantation profile: as that no significant dependence was observed, all subsequent
implantations ($\textrm{4th}$ to $\textrm{7th}$) were performed with a beam of 23 keV.
}
\begin{tabular}
[c]{ccccc}\hline\hline
Irradiation & Energy & Peak of depth profile\footnotemark[1] & Fluence &
Accumulated Fluence\\
\multicolumn{1}{l}{session \#} & (keV) & (nm) & (10$^{16}$\thinspace
ions/cm$^{2}$) & (10$^{16}$\thinspace ions/cm$^{2}$)\\\hline
1 & 23 & 50 & 0.42 & 0.42\\
2 & 10 & 23 & 0.34 & 0.76\\
3 & 30 & 64 & 0.50 & 1.26\\
4 & 23 & 50 & 1.21 & 2.47\\
5 & 23 & 50 & 1.25 & 3.72\\
6 & 23 & 50 & 1.77 & 5.49\\
7 & 23 & 50 & 0.97 & 6.46\\\hline\hline
\end{tabular}
\footnotetext[1]{Estimated using the \$%
$\backslash$%
textrm\{SRIM\}\$ code.\cite{Ziegler10-SRIM}}\label{TabI-Fluence-SRIM}%
\end{table*}%

\section{Results}%

\begin{figure}[th]%
\centering
\includegraphics[
height=6.776cm,
width=7.6484cm
]%
{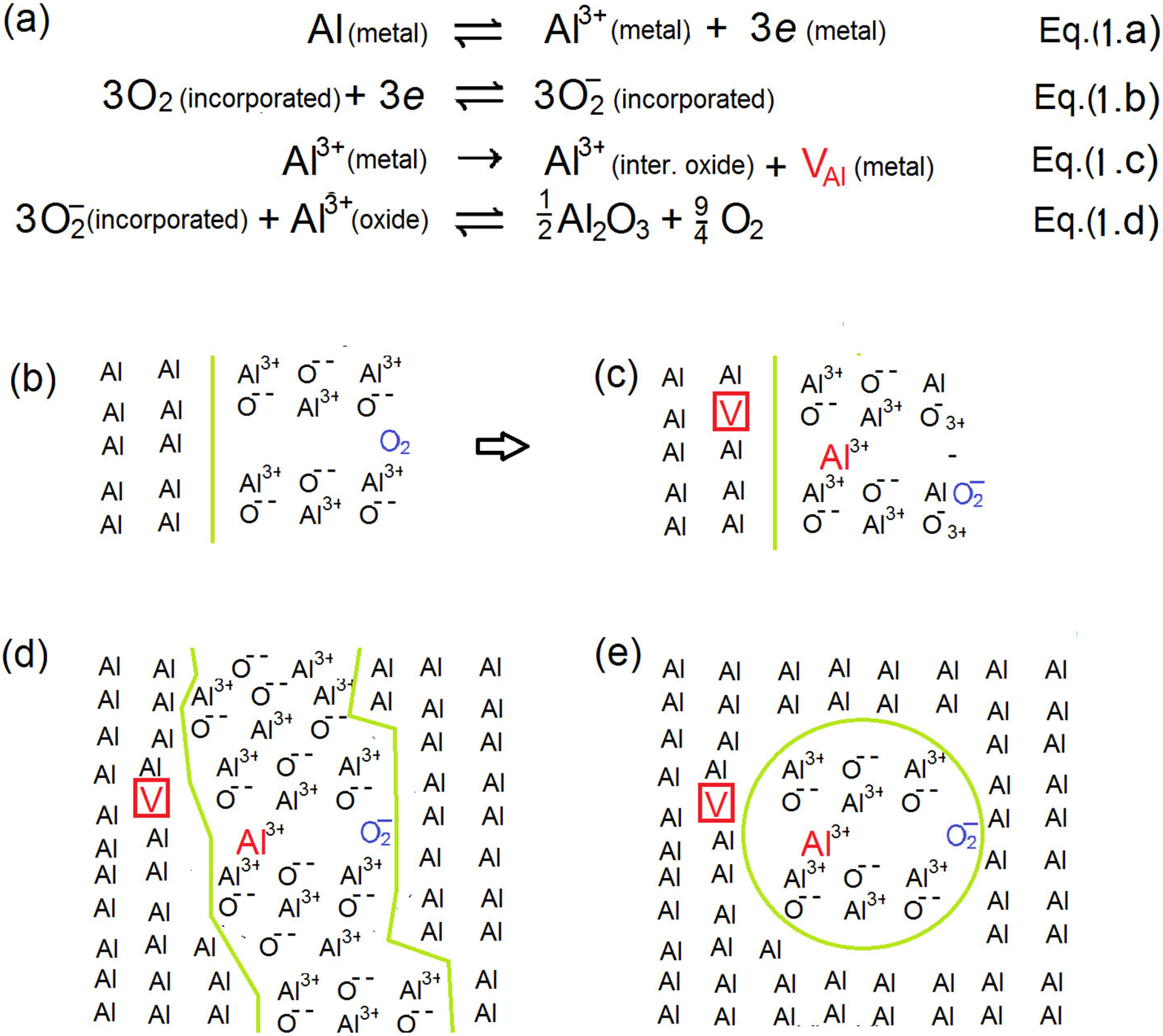}%
\caption{(a) A multistep kinetic scheme for oxidation of aluminum as described
in the Cabrera-Mott model (adapted from Ref.{\thinspace
\onlinecite{Boggio66-Metal-ThinFilms-Theory})}. Directly after ionization
[Eq.{\thinspace(}1.a)], electrons pass freely through the oxide interface
until reaching incorporated oxygen which are then ionized to O$_{2}^{-}$
[Eq.{\thinspace(}1.b)]. This charge transfer together with the left-behind
positive Al$^{3+}$ induce a local electric field which drives the slow
migration of Al$^{3+}$ across the oxide interface leaving behind a vacancy
\color{red}V$_{\text{Al}}$ \color{black} [Eq.{\thinspace(}1.c)].
Equation{\thinspace(}1.d) indicates a typical formation of Al$_{2}$O$_{3}$ by
the combination of migrated Al$^{3+}$ and ionized O$_{2}^{-}$. Panels (b) and
(c): Illustration of the reaction at metal-oxide interface (represented by the
solid green line) of a directly exposed film [panel (b) is before an event of
Eq.(1.a) while panel (c) is after the event of Eq. (1.c); adapted from
Ref.{\thinspace\onlinecite{Atkinson85-Growth-Oxides-Films}}]. (d) We consider
that, during the codeposition process of a granular film, the incorporated
oxygen does not enter as an idle and neutral entity, rather it does react with
Al matrix, just as in the normal oxidation process of Cabrera-Mott leading to
an interaction similar to that described in panels (a)-(c). Ultimately this
accumulates into an extended \ nanosized grain. (e) Similar reaction occurring
at the boundary of an isolated oxide grain of an irradiated film. In all
cases, incorporated and ionized oxygen are represented, with no loss of
generality, by O$_{2}$ and O$_{2}^{-}$. }%
\label{Fig1-Al-Kinetics-Cabrera-Mott}%
\end{figure}
\begin{figure}[tb]%
\centering
\includegraphics[
height=6.0313cm,
width=7.6484cm
]%
{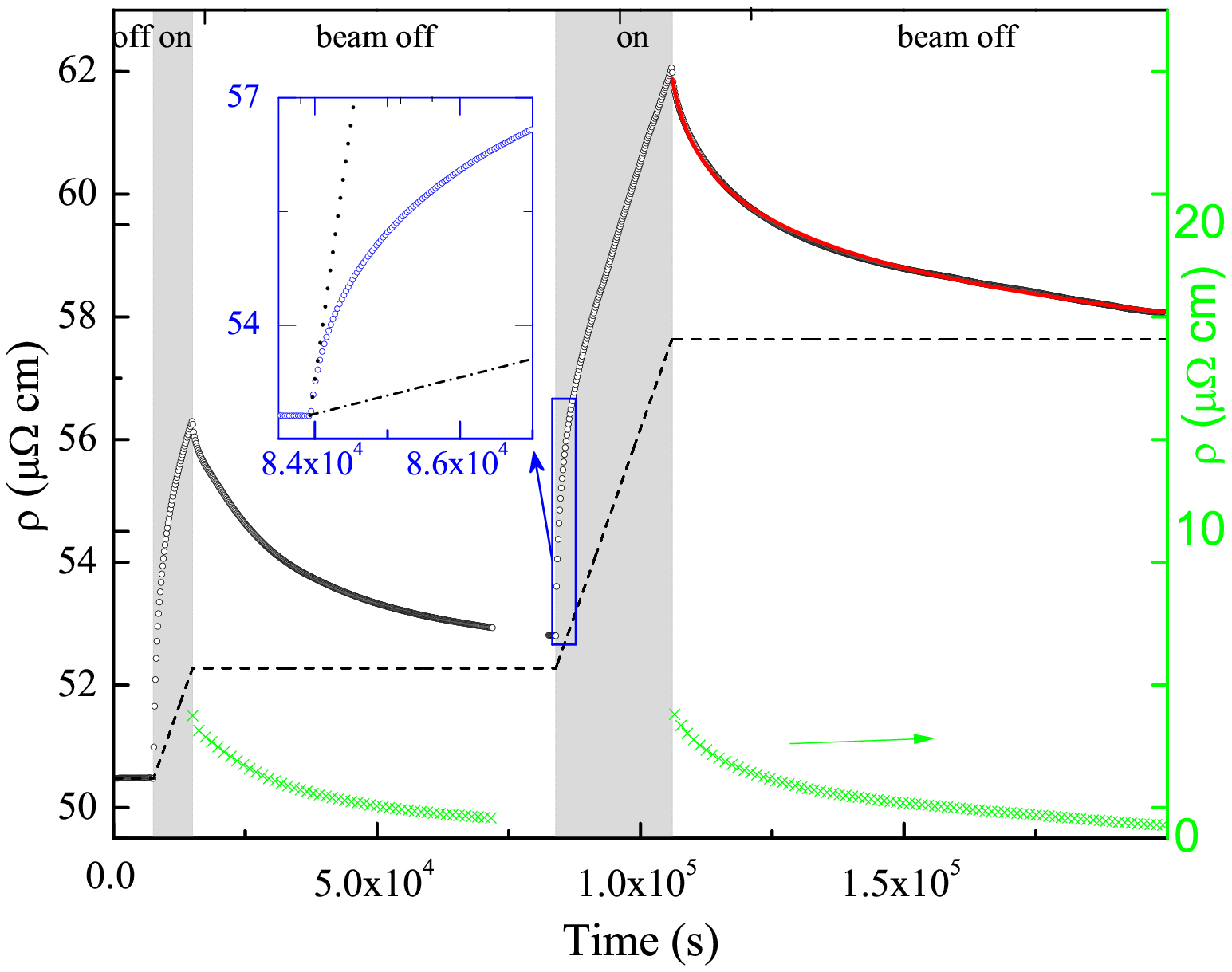}%
\caption{$\rho(t$, 300K,$6\mathrm{th})$ measured with a current density
$j$=10$^{7}$A/m$^{2}$ during two consecutive O-irradiation\ sessions without
breaking a vacuum of 1.0$\times10^{-7}$ Torr. The irradiation intervals are
shown as vertical gray areas. The inset\textit{ }highlights the presence of
relaxation processes: In their absence, the resistivity rise should have
followed the dotted line. During the steady state, the resistivity rises
linearly with a rate given by the dot-dashed line in the inset. Directly after
beam stoppage, there are two contributions: a fast-decaying one governed by
one class of relaxations (depicted as green crosses with values given by the
right-hand y axis) and a slow contribution governed by a second class of
relaxations; the latter is evident as a weak decay in the almost horizontal,
dashed lines. The fast-decaying $\rho_{d}(t>106000$s, 300K, $6\mathrm{th})$
is\ well fitted to Eq.(5) of the first of Refs.
{\onlinecite{Dorey69-Resistivity-vs-time,*Day95-resistivity-GrainSize-Ti-Films}}
(solid red line).}%
\label{Fig2-Al-RvsTime-Anneal}%
\end{figure}
\begin{figure}[tb]%
\centering
\includegraphics[
height=12.8052cm,
width=7.6484cm
]%
{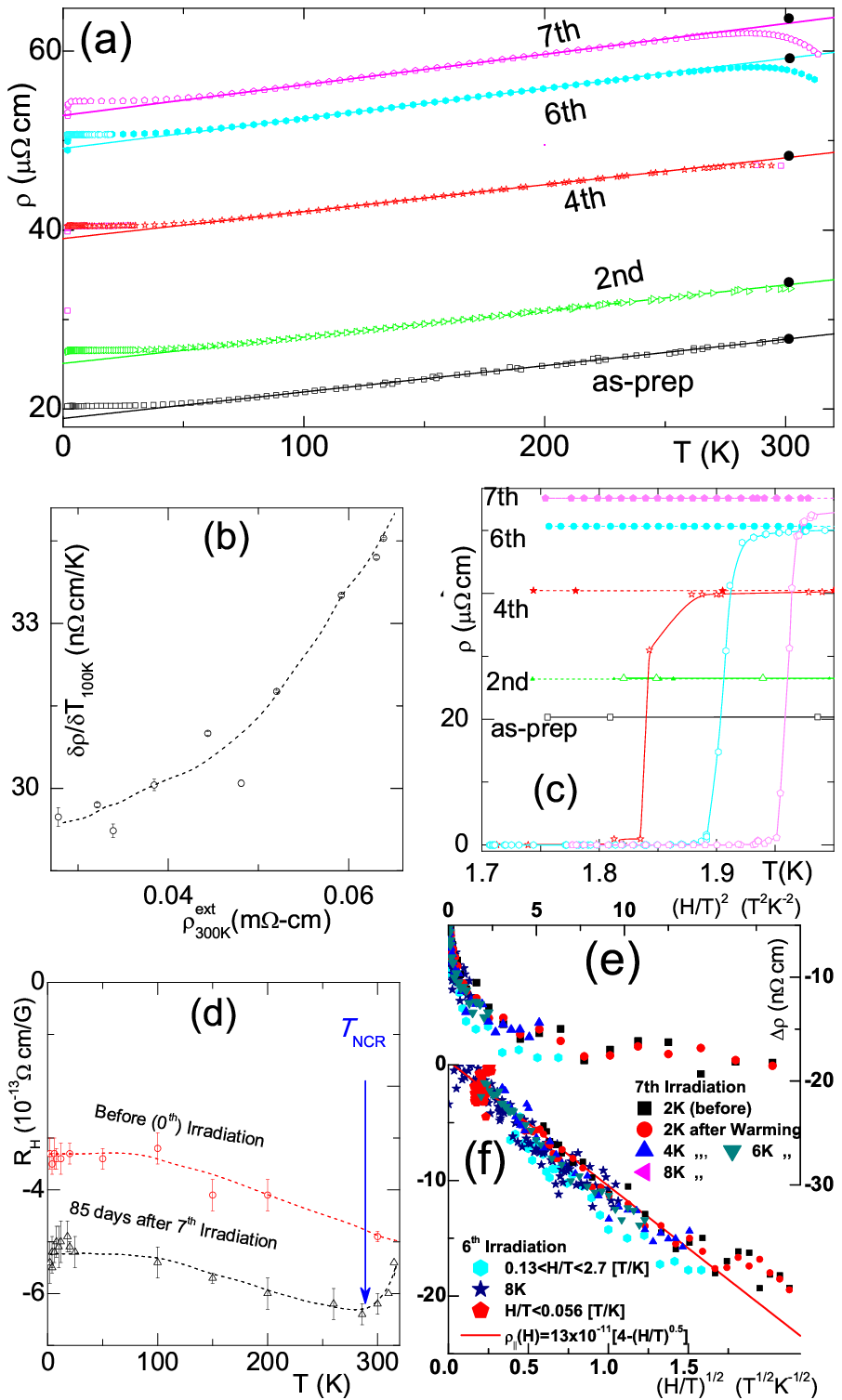}%
\caption{(\textit{a}) Representative $\rho(T,n\mathrm{th})$ curves. The solid
lines are linear fit with a slope $\left(  \partial\rho/\partial T\right)
_{\text{100-220K}}$. Due to aging effects, a unique room-temperature
resistivity for each warming-up measuring cycle is taken to be the
extrapolated $\rho_{\text{300K}}^{ext}$, solid circle on the high-$T$ linear
extrapolation. (\textit{b}) Evolution of $\left(  \partial\rho/\partial
T\right)  _{\text{100-220K}}$ as a function of $\rho_{\text{300K}}^{ext}$.
(\textit{c}) An expanded view of low-$T$ $\rho(T,H$, $n\mathrm{th})$ curves of
panel \textit{a}. \ Open (solid) symbols denote zero-field (5kOe) curve.
(\textit{d}) Hall coefficients of as-prepared film and that of the same film
measured 85 days after the 7$^{th}$ irradiation. (\textit{e}) $\Delta\rho
(H/T$, $n\mathrm{th})=\rho(H/T$, $n\mathrm{th})-\rho(0,T$, $n\mathrm{th})$
$versus$ \ $\left(  H/T\right)  ^{2}$: showing the breakdown of the $\left(
H/T\right)  ^{2}$ scaling. These curves are from $n\mathrm{th}=$6, 7
irradiation and are limited to the range of $\ H_{c2}<H<$20\thinspace kOe and
$T_{c}<T<T_{\text{K}}^{mn}$ (small $\nicefrac{H}{T}$). (\textit{f}) The same
as panel \textit{e} but are scaled to $\sqrt{\nicefrac{H}{T}}$. This better
scaling is emphasized by the solid line fit $\Delta\rho(\nicefrac{H}{T}$,
6$\mathrm{th})\propto\left(  H/T\right)  ^{\nicefrac{1}{2}}$. }%
\label{Fig3-Al-RvsT-Iraddiated}%
\end{figure}
\begin{figure}[h]%
\centering
\includegraphics[
height=9.7676cm,
width=7.6484cm
]%
{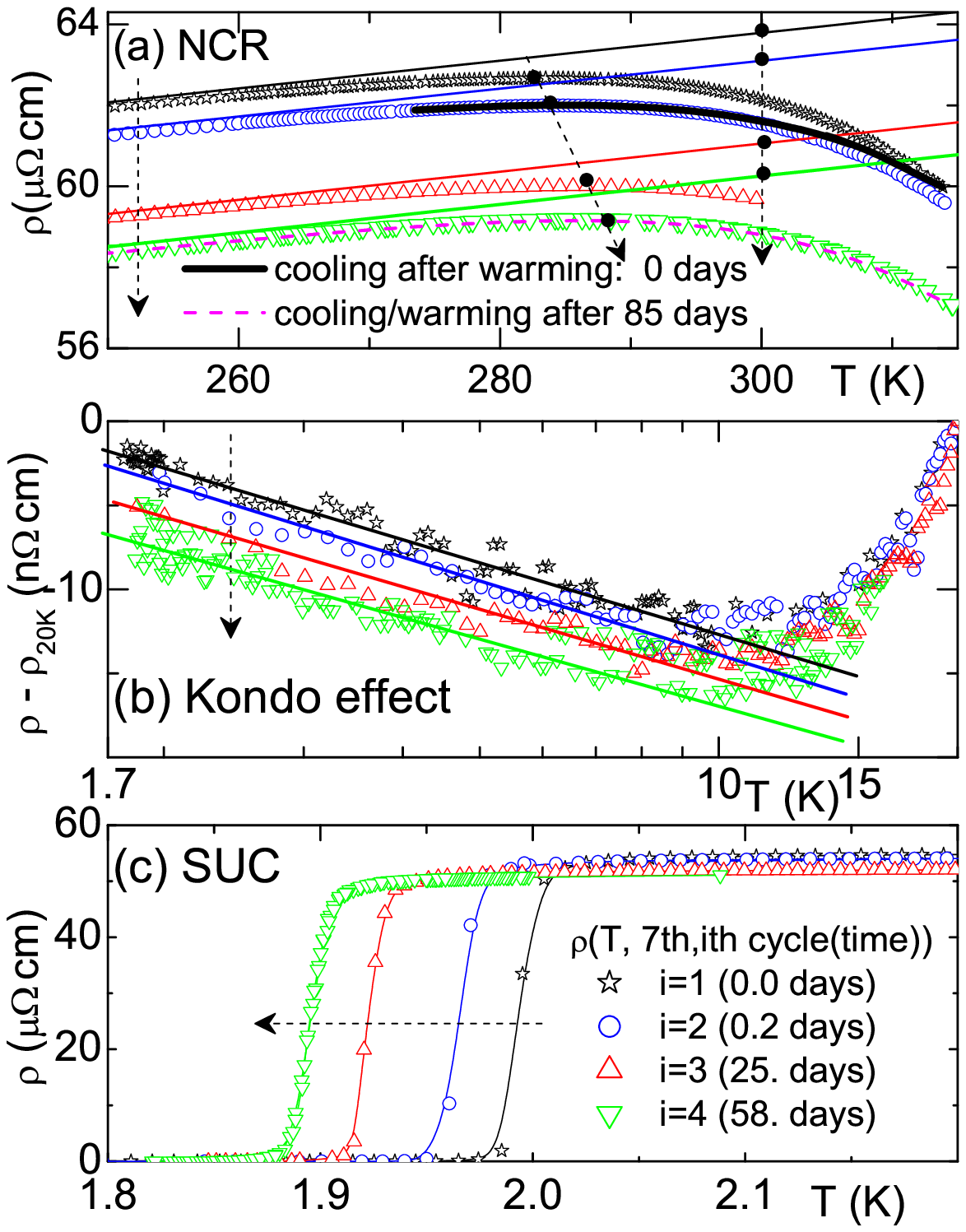}%
\caption{Manifestation of the influence of aging on the three main feature of
the phase diagrams. The $i\mathrm{th}$ cycle represents the order and time (in
days) of the cooling-warming measurement carried out after the 7$\mathrm{th}$
irradiation session. (a) The influence of aging on the position and height of
the peak maximum\ of NCR. The thick black line is the cooling curve directly
after the first, $i=1$, warming-up measurement while the dashed line is the
cooling and warming curves of $i=5$ cycle after 85 days. The solid thin
straight lines are linear fits within the range 100$\leq T\leq$220\thinspace
K. The solid circles represent each of $\rho(T_{\text{NCR}},$%
0kOe$,7\mathrm{th},i\mathrm{th})$ and $\rho_{\text{300\thinspace K}}^{ext}%
($0kOe$,7\mathrm{th},i\mathrm{th})$. (b) Thermal evolution of $\rho
(T,$5\thinspace kOe$,7\mathrm{th},i\mathrm{th})-$ $\rho($20\thinspace
K$,$0\thinspace kOe$,7\mathrm{th},i\mathrm{th})$ in a semilog plot. For
$T<$10\thinspace K, this shows a \textit{log-in-T} dependence which is not a
quantum localization effect since our film does not manifest a 2-dimensional
character. (c) The degradation of $T_{\text{C}}$ with aging. The arrows
highlight the tendency of aging influence while the solid lines in panels (b)
and (c) are guides to the eye. Just as for $T_{\text{NCR}}$ of panel (a),
there is no appreciable change in $T_{\text{K}}^{\min}$ and $T_{\text{c}}$ for
$t>85$ days.}%
\label{Fig4-Al-Aging-7th-Iraddiation}%
\end{figure}

\subsection{The two classes of relaxation processes}

Oxygen implantation is a violent process that leads to an unstable state. A
drive towards equilibrium requires an activation of some relaxation channels.
Different from neutron or electron
irradiation,\cite{Isebeck66-Reconvery-in-Al-NeutronIrradiation,*Sosin63-Aluminum-Recovery-AfterIrradiation,*DeSorbo59-Kinetics-Vacancy-Al,*Khellaf02-Quench-Aluminum}
O irradiation provides additional process(es) related to the Cabrera-Mott
oxidation process\cite{Cabrera49-oxidation-ThinFims} (see Fig.
\ref{Fig1-Al-Kinetics-Cabrera-Mott}). Therefore, one expects (see Fig.
\thinspace\ref{Fig2-Al-RvsTime-Anneal}) two classes of relaxation processes.
The first class consists of conventional, relatively fast, relaxation
processes,\cite{Isebeck66-Reconvery-in-Al-NeutronIrradiation,*Sosin63-Aluminum-Recovery-AfterIrradiation,*DeSorbo59-Kinetics-Vacancy-Al,*Khellaf02-Quench-Aluminum}
which are manifested, for pure Al, as three recovery
stages:\cite{Sosin63-Aluminum-Recovery-AfterIrradiation} Stage I is dominant
below $\sim$50\thinspace K; stage II is centered at $\sim$140\thinspace K,
while stage III operates within 190$<T<$250\thinspace K. The second class
consists of much slower (in weeks)\ relaxation processes and are dominant at
sufficient O-implantation level and higher temperatures, $T>$250\thinspace K.
As can be seen in Eqs. 1(a) -- 1(d) of Fig.\ref{Fig1-Al-Kinetics-Cabrera-Mott}%
, the overall process involves charge transfer, charge trapping, vacancies
creation, defect migrations and annihilation,
etc.\cite{Atkinson85-Growth-Oxides-Films} We show below that the processes
associated with the second class are the driving factors behind the
above-mentioned modifications in the phase diagram.

It is worth mentioning that (i) the reaction involving Eqs.(1a)-1(d) continues
until\ equilibrium is established via opposing forces related to diffusion and
disassociation. (ii) Oxidation in both the granular and irradiated films, in
contrast to the exposed case, occurs within the film bulk/profile depth and,
furthermore, the involved quantity of oxygen is fixed after ending the
incorporation process. (iii) For the particular case of irradiated film, the
initial uniformly distributed out-of-equibrium state relaxes back to
equilibrium via the same two relaxation processes. As far as the incorporated
oxygen is concerned, its reaction is similar to the oxidation process
described by Eqs. (1a)-1(d). Ultimately, this leads to an incipient
germination of centres of Al$_{2}$O$_{3}$ which with further kinetics
accumulates into a nano-sized bubble: The limitation into nanosized grains is
identical to the case of nano-sized Al$_{2}$O$_{3}$ surface in conventional
oxidation of metallic Al.

\subsection{The negative-curvature resistivity (NCR)}

Resistivity curves $\rho(1.7\leq T\leq315$\thinspace K, $n\mathrm{th})$ of
Figs.\thinspace\ref{Fig3-Al-RvsT-Iraddiated}(a) indicate conclusively that the
surge of $T_{\text{c}}(x)$ enhancement [Fig.\ref{Fig3-Al-RvsT-Iraddiated}(c)],
the Kondo behavior [Figs. \ref{Fig3-Al-RvsT-Iraddiated}(e) and
\ref{Fig4-Al-Aging-7th-Iraddiation}(b)], as well as the NCR effect [Fig.
\ref{Fig4-Al-Aging-7th-Iraddiation}(a)] are correlated with each other and
that all are much accentuated with each subsequent $n\mathrm{th}$ irradiation
[Fig. \ref{Fig3-Al-RvsT-Iraddiated}(a)]. These features were not observed in
conventional recovery stages; accordingly they must be associated with the
second class of relaxation processes. In addition, there are aging effects
that are much accentuated for $T>T_{\text{NCR}}$; in contrast, there are no
signs of aging in $\rho(t,T<T_{\text{NCR}},$ $7\mathrm{th})$ even when the
film is repeatedly recycled within $T<T_{\text{NCR}}$. The manifestation of a
slow (in weeks) aging of both the position of $T_{\text{NCR}}$ and the
magnitude of $\rho(T_{\text{NCR}})$ rules out any interpretation in terms of
crystalline electric field splitting. It is worth adding that a similar aging
effect had not been explicitly studied in Al granular film and that
linear-in-$T$ behavior for granular films were observed only for samples with
$\rho_{\text{300K}}$%
$<$
100$\mu\Omega$-cm.

After three months, $\rho(T,$ $7\mathrm{th})$ traverses the NCR peak
reversibly and with no hysteresis (though with a\ lower magnitude and a higher
$T_{\text{NCR}}$): This nonhysteresis feature can be identified in the
reported results of granular Al
films.\cite{Bachar15-Mott-granular-Al,*Bachar13-Kondo-granular-Al,Bachar14-PhD-Thesis}%
{ The similarity in both the reported and our }$T_{\text{NCR}}${ (see plot of
Fig.\thinspace\ref{Fig5-Al-granular-PhaseDiagram}) reveals a similarity in the
mechanism behind this NCR}.

{Figure\thinspace\ref{Fig3-Al-RvsT-Iraddiated}(d) indicates that the}
conductivity enhancement ($T>T_{\text{NCR}}$) occurs concomitantly with a
decrease in the magnitude of Hall coefficient. We associate the NCR event to a
thermally assisted liberation of trapped charges (with a binding potential of
$\sim k_{B}T_{\text{NCR}}$). Such trapped charges are possibly electrons
(neighboring an anionic vacancy) or holes or hole pairs (trapped near an
oxygen ion which is adjacent to Al$^{3+}$
vacancy).\cite{Gamble64-EPR-Al2P3,*Lee76-holes-Al2O3} After reaching a steady
state, such a disassociation-association process is reversible leading to a
reproducible NCR peak at $T_{\text{NCR}}$. It is worth mentioning that studies
on $\gamma$-irradiated Al$_{2}$O$_{3}$ identified a thermally assisted
dissociation of hole pairs into single hole at 384\thinspace K and an
annihilation of holes at 533\thinspace
K.\cite{Gamble64-EPR-Al2P3,*Lee76-holes-Al2O3}

\subsection{The Kondo behavior}

On lowering the temperature much below $T_{\text{K}}^{\min}$, we observed a
\textit{log-in-T} feature [Fig. \ref{Fig4-Al-Aging-7th-Iraddiation}(b)] as
well as a negative magnetoresistivity [Figs. \ref{Fig3-Al-RvsT-Iraddiated}(e)
and 3(f)] that scales with $\sqrt{\nicefrac{H}{T}}$ rather than the expected
$\left(  \nicefrac{H}{T}\right)  ^{2}$. Same $\sqrt{\nicefrac{H}{T}}$ scaling
can be observed in magnetoresistivity of granular films with $\rho
_{\text{300K}}$%
$<$
100$\mu\Omega$-cm (i.e. within a range similar to
ours).\cite{Bachar14-PhD-Thesis} As that we did not observe a $T^{-3/2}$
dependence, that the \textit{log-in-T} contribution and $T_{K}^{\min}$ occur
at a relatively high-$T$ range, and that our films are considered to be 3$d$,
we do not associate this scaling to a localization\ stemming from 2$d$ quantum
corrections. More compelling evidence of a Kondo-like behavior is the direct
observation, by Bachar \textit{et al.}%
\cite{Bachar15-Mott-granular-Al,*Bachar13-Kondo-granular-Al,Bachar14-PhD-Thesis}%
, of free spins in oxygen-incorporated granular Al films{. }Similar to the NCR
case, the Kondo behavior become more accentuated on subsequent $n\mathrm{th}$
irradiation [but degraded by aging as in Fig.\thinspace
\ref{Fig4-Al-Aging-7th-Iraddiation}(b)].%

\begin{figure*}[tbp] \centering
\raisebox{-0cm}{\includegraphics[
height=6.7656cm,
width=12.0946cm
]%
{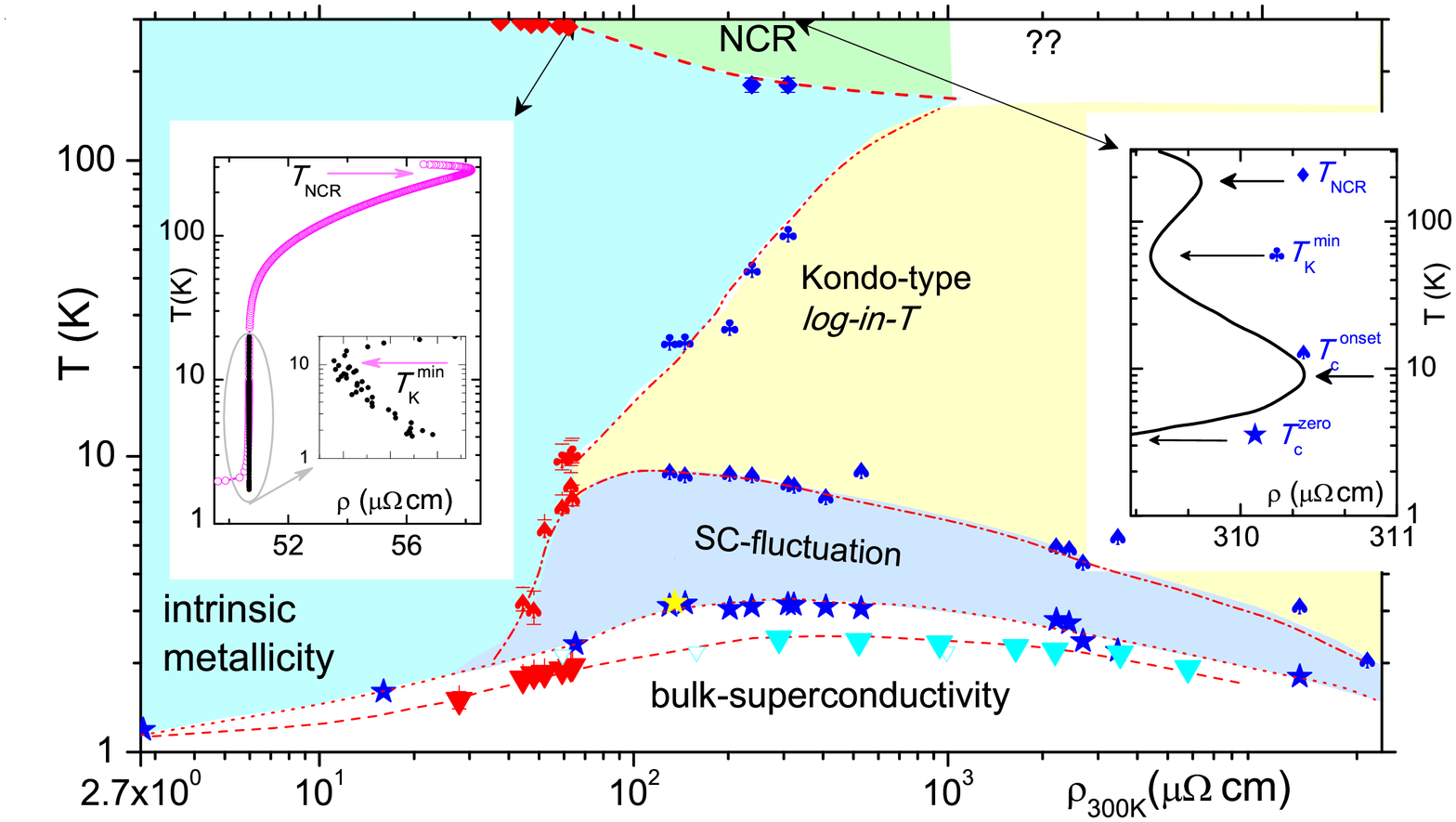}%
}
\caption{Normal-state and superconducting phase diagram of different Al thin
films shown as a log-log plot of $T$ vs $\rho_{300\text{K}}$.
\color{red}$\blacktriangledown$, $\spadesuit$,
$\clubsuit,\blacklozenge$\color{black}: $T_{\text{c}}^{zero}$, $T_{\text{c}}^{onset}(=$ $T_{\text{K}}^{\max})$, $T_{\text{K}}^{\min}$, $T_{\text{NCR}}$,
resp., as obtained by this work. \color{blue}$\bigstar$, $\spadesuit$,
$\clubsuit$, $\blacklozenge$\color{black}: $T_{\text{c}}^{zero}$,
$T_{\text{c}}^{onset}(=$ $T_{\text{K}}^{\max})$, $T_{\text{K}}^{\min}$,
$T_{\text{NCR}}$, resp., of{ granular films }deposited at 77 K
(Refs. {\thinspace
\onlinecite{Bachar15-Mott-granular-Al,*Bachar13-Kondo-granular-Al,Bachar14-PhD-Thesis}}){. }T{he location of each thermal event is explicitly shown at the}
right-side inset.{ }\color{cyan}$\triangledown,\blacktriangledown
$\color{black}: $T_{c}$ of granular films deposited at room-temperature
(Ref.{\thinspace}\onlinecite{Abeles67-Hc-granular-SUCs} and
{\onlinecite{Chui81-Nano-Al-MagRes-Localizatio,*Chui81-Tc-Hc-Granular-Al-SUC,*Mui84-Granular-Al-MagRes}}, resp.). \color{yellow}$\bigstar$\color{black}: $T_{c}$\ after O-implantation
at liquid helium (Ref.{\thinspace
\onlinecite{Lamoise75-Irradiation-Al-Resistivity-annealing})}.
Left-hand-side inset: A semilog plot of our irradiated-film's $\rho(T,6^{th},$0 kOe$)$ and
$\rho(T,6^{th},$5 kOe$)$ (open and close circles, resp.). Right-hand-side inset:
A semilog plot of $\rho(T)$ curve of a granular sample having
$\rho_{300K}=$310 $\mu\Omega$-cm (taken from Refs.{\thinspace
\onlinecite{Bachar15-Mott-granular-Al,*Bachar13-Kondo-granular-Al,Bachar14-PhD-Thesis}}). The main graph is extrapolated down to 2.75$\mu\Omega$-cm of bulk
Al.\cite{Desai84-Al-resistivity} The lines are guides to the eye
}\label{Fig5-Al-granular-PhaseDiagram}%
\end{figure*}%

{We associate this Kondo contribution to scattering off} irradiation-induced
paramagnetic,\ color-center-type, defects located at the interface between Al
and Al$_{2}$O$_{3}$ grains. It is recalled that\ magnetic defects located at
the Al-Al$_{2}$O$_{3}$ interface had been suggested by various workers who
studied the 1/$f$ flux noise in Al-based
SQUID.\cite{Lee14-Mag-Al2O3-Al-interface,*Anton13-SQUID-Noise,*Bluhm09-MagSuc-Al-Film,*Sendelbach08-Maggnetism-in-SQUID}
Moreover, such magnetic centers were observed in irradiated Al$_{2}$O$_{3}$ by
electron spin resonance studies and were attributed to either unpaired
electrons which are trapped at an anion vacancy or to a hole trapped near an
oxygen ion which is adjacent to Al$^{3+}$
vacancy.\cite{Gamble64-EPR-Al2P3,*Lee76-holes-Al2O3} On assuming that
implantation has introduced
complexes\cite{Linker76-Tc-Mo-Implantation,*Linker82-Implantation-Mo-Tc-Enhancement}
(e.g., Al$_{2}$O$_{3}$), which are capable of stabilizing these centers (see
Fig. \ref{Fig1-Al-Kinetics-Cabrera-Mott}), one is able to explain the
following: (i) Kondo behavior is not observed in pure Al films, not even when
these are irradiated with nonchemically active projectiles. (ii) Kondo
behavior is absent at the lower left-hand side of Fig.\thinspace
\ref{Fig5-Al-granular-PhaseDiagram}. Here the metallic and screening character
are sufficiently strong to oppose the formation of paramagnetic centers.
Finally, (iii) the surge, operation and manifestation of a Kondo process is
similar in both the granular and irradiated films: Indeed, a plot of the
obtained minimum point of resistivity, $T_{\text{K}}^{\text{min}}$, versus
$\rho_{\text{300K}}^{ext}$ in Fig.\thinspace
\ref{Fig5-Al-granular-PhaseDiagram} evolves smoothly and extrapolates directly
into a $T_{\text{K}}^{\text{min}}-$ $\rho_{\text{300K}}$ curve taken from
Refs.
{\onlinecite{Bachar15-Mott-granular-Al,*Bachar13-Kondo-granular-Al,Bachar14-PhD-Thesis}}%
.

\subsection{The enhancement of $T_{\text{c}}$}

Figure \ref{Fig3-Al-RvsT-Iraddiated}(c) illustrates the accentuation of
$T_{\text{c}}$ enhancement with the subsequent $n\mathrm{th}$ irradiation:
Note that (i) the superconducting transitions width ($\Delta T_{c}%
^{10-90\%}<80\,$mK) is sharp and that (ii) superconductivity is quenched on an
application of $H>H_{c2}$($T,n\mathrm{th})\approx$5\thinspace kOe.
Figure\thinspace\ref{Fig4-Al-Aging-7th-Iraddiation}(c), on the other hand,
demonstrates $T_{\text{c}}$ decrease with aging.

Figure \ref{Fig5-Al-granular-PhaseDiagram} summaries the $T_{\text{c}}$
enhancement as obtained from this work [see Figs.\thinspace
\ref{Fig3-Al-RvsT-Iraddiated}(c)] as well as those reported on
granular\cite{Abeles75-Granular-Films-Review,Deutscher08-Nano-Granular-SUC-Review,Bachar15-Mott-granular-Al,*Bachar13-Kondo-granular-Al,Bachar14-PhD-Thesis}
and irradiated
films.\cite{Lamoise75-Irradiation-Al-SUC,*Lamoise76-Enhanced-SUC-Al-Alloys,Lamoise75-Irradiation-Al-Resistivity-annealing,Meunier77-Irradiation-Al-Percolation-SUC}
It is remarkable that all data, new and old, follow the same phase boundaries
across the whole available region: This finding is far from being trivial.

Evidently, $T_{\text{c}}$ is enhanced monotonically at lower oxygen
incorporation, passes through a maximum and afterwards decreases
monotonically. Numerous theoretical models were suggested for the explanation
of $T_{c}$ enhancement [see, e.g., Refs.\thinspace
{\onlinecite{Abeles75-Granular-Films-Review,Beloborodov07-Transport-Granular-Review,Deutscher08-Nano-Granular-SUC-Review,Garland68-Disorder-enhanced-Tc-Al,*Klein68-Phonon-Granular-Al,*LEGER69-Enhanced-SUC,Allender73-Excitonic-SUC,Kresin08-Tc-enhanced0Nanoclusters,*Bose10-SUC-Sn-Nanocluster,*Mayoh14-SUC-NanoGranular}}%
]. One of these relates $T_{c}$ enhancement to quantum-size effects in shell
structures\cite{Kresin08-Tc-enhanced0Nanoclusters,*Bose10-SUC-Sn-Nanocluster,*Mayoh14-SUC-NanoGranular}
but various studies, including this work, indicate that grain-size character
(spatial confinement) is not a decisive factor in $T_{c}$ enhancement.
Magnetic-based mechanisms can also be ruled out since $T_{c}$ enhancement in
Fig.\thinspace\ref{Fig5-Al-granular-PhaseDiagram} occurs much earlier than the
region wherein possible magnetic fluctuation, if any, is expected. The BCS
mechanism,\cite{Garland68-Disorder-enhanced-Tc-Al,*Klein68-Phonon-Granular-Al,*LEGER69-Enhanced-SUC}
on the other hand, can not be excluded since it describes successfully the
superconductivity of elemental Al and, by extension, the lower limit of this
phase diagram (left-hand side of $T_{c}$ dome of Fig.
\ref{Fig5-Al-granular-PhaseDiagram}). Then, using a weak-limit BCS
approximation, one expects to identify the essential ingredient behind $T_{c}$
enhancement. A variation in $\partial T_{c}(\partial x)$ can be expressed in
terms of a sum of a variation in Debye temperature $\theta_{D}$, in the
pairing potential $V$ arising from electron-phonon coupling $\lambda$, and in
the density of states $N_{F}$ at Fermi energy:
\begin{equation}
\frac{\partial\ln T_{c}}{\partial x}=\frac{\partial\ln\theta_{D}}{\partial
x}+\frac{1}{VN_{F}}\left(  \frac{\partial\ln V}{\partial x}+\frac{\partial\ln
N_{F}}{\partial x}\right)  \tag{2}\label{Eq-BCS-Tc-Disorder}%
\end{equation}
The Hall effect\cite{Bandyopadhyay82-HallEffect-granular-Al} and specific
heat\cite{Greene72-Cp-Granular-Al} measurements on granular films indicated
that $N_{F}$ decreases with $x$. Our Hall curves in
Fig.\ref{Fig3-Al-RvsT-Iraddiated}(d) do confirm this trend: While $R_{H}$ of
as-prepared film is the same as -3.4$\times10^{-13}$ $\Omega$cm/G of pure Al,
that of irradiated one is more negative and exhibits an upturn above
$T_{\text{NCR}}$. $\theta_{D}$, just as well decrease with $x$%
.\cite{Greene72-Cp-Granular-Al} Then, $T_{c}$ enhancement must be related to
an increase in $V$ which, in turn,\ is related to $\lambda$%
.\cite{McMillan-Tc-StrongCoupling} Indeed, Fig.\thinspace
\ref{Fig3-Al-RvsT-Iraddiated}(b) confirms this increase in $\lambda$ by
demonstrating an increase in the metallic slope\cite{Allen86-transport-metals}
$\left(  \partial\rho/\partial T\right)  _{\text{100-220\thinspace K}}$ with
$x$. In fact it is almost \emph{three times} higher than the 12\thinspace
n$\Omega$-cmK$^{-1}$ reported for bulk Al.\cite{Desai84-Al-resistivity}

We associate such an increase in $\lambda$, $V$, and $T_{\text{c}}$
enhancement\cite{Garland68-Disorder-enhanced-Tc-Al,*Klein68-Phonon-Granular-Al,*LEGER69-Enhanced-SUC,MATTHIAS80-Enhanced-Tc-YxIr}
to a softening of the lattice, which is facilitated by the presence of
vacancies that are created and stabilized during the\ "\textit{oxidation-like}%
" process. Thus aging of $T_{\text{c}}$ is driven by partial removal of these
softening-inducing defects (by recombination, sinking,
etc.\cite{Atkinson85-Growth-Oxides-Films}). The presence of such an aging
process explains the manifestation of\ two $T_{c}(x)$ branches in
Fig.\thinspace\ref{Fig5-Al-granular-PhaseDiagram}: One branch is associated
with $T_{c}^{>T_{\text{NCR}}}$ of films deposited or irradiated above
$T_{\text{NCR}}$,
\cite{Abeles67-Hc-granular-SUCs,Chui81-Nano-Al-MagRes-Localizatio,*Chui81-Tc-Hc-Granular-Al-SUC,*Mui84-Granular-Al-MagRes}
while the other with $T_{c}^{<T_{\text{NCR}}}(x)$ of films
deposited/irradiated below $T_{\text{NCR}}$%
.\cite{Bachar15-Mott-granular-Al,*Bachar13-Kondo-granular-Al,Bachar14-PhD-Thesis,Lamoise75-Irradiation-Al-Resistivity-annealing}
Evidently~aging effects lead to $T_{c}^{>T_{\text{NCR}}}(x)<T_{c}%
^{<T_{\text{NCR}}}(x)$; this is also evident in that the evolution of our
$T_{c}^{zero}(x)${ (prepared and irradiated at 300K) }is in excellent
agreement with that of $T_{\text{c}}^{>T_{\text{NCR}}}(x)$%
.\cite{Abeles67-Hc-granular-SUCs,Chui81-Nano-Al-MagRes-Localizatio,*Chui81-Tc-Hc-Granular-Al-SUC,*Mui84-Granular-Al-MagRes}
Remarkably, each of $T_{c}^{>T_{\text{NCR}}}(x)$ and $T_{c}^{<T_{\text{NCR}}%
}(x)$ follows a dome-like
evolution{\cite{Deutscher73-Granular-SUC-Films,*Deutscher73-Granular-SUC-zeroDimension}%
} when plotted on a log-log scale; the maximum is attained at 2.3K
(Refs.\thinspace
{\onlinecite{Chui81-Nano-Al-MagRes-Localizatio,*Chui81-Tc-Hc-Granular-Al-SUC,*Mui84-Granular-Al-MagRes})}
for the former while at 3.2 K (Ref.\thinspace
{\onlinecite{Bachar15-Mott-granular-Al,*Bachar13-Kondo-granular-Al,Bachar14-PhD-Thesis})}
for the latter.

\section{Further discussion and summary}

Impurities in an O-irradiated Al film consist mainly of implantation-induced
chemical
complexes.\cite{Linker76-Tc-Mo-Implantation,*Linker82-Implantation-Mo-Tc-Enhancement}
Then, most of impurity-stabilized defects should be located at the border
between, e.g., Al$_{2}$O$_{3}$ and metallic grains. These impurity-stabilized
defects are assumed to consist of paramagnetic centers, trapped electrons,
trapped holes or hole pairs, or vacancies (see
Fig.\ref{Fig1-Al-Kinetics-Cabrera-Mott}). A reduction of these defects by any
recombination or annihilation process would lead to (i) a reduction of
scattering centres (aging of $\rho_{\text{300K}}$), (ii) a reduction of the
thermally assisted liberation or annihilation of trapped charges above
$T_{\text{NCR}}$ (aging of NCR), (iii) a reduction of the paramagnetic centers
(lowering of $T_{\text{K}}$) and (iv) a reduction in lattice softening
(degradation of $T_{c}$).

It is notable that the surge of Kondo behavior occurs just below the dome
maximum of Fig.\thinspace\ref{Fig5-Al-granular-PhaseDiagram}
[Ref.{\onlinecite{Bachar15-Mott-granular-Al,*Bachar13-Kondo-granular-Al,*Pracht15-SUC-dome-Granular-Al,Bachar14-PhD-Thesis,Deutscher73-Granular-SUC-Films,*Deutscher73-Granular-SUC-zeroDimension}]
and that} the monotonic evolution of Kondo effect is accompanied by a slowing
down, leveling out, and eventual decay of $T_{c}(x)$. Accordingly, the dome
like evolution of $T_{c}(x)$ is attributed to a compromise between an
enhancement\ trend (due to lattice softening) and a suppression trend (due to
Abrikosov-Gorkov pair-breaking process). As $T\rightarrow$ $T_{\text{c}}^{+}$
($T<$ $T_{\text{K}}^{\min}$), a competition between\ spin-flip scattering and
Cooper pairing leads to a downward deviation away from the \textit{log-in-T}
behavior and as such to an eventual resistivity maximum at $T_{\text{K}}%
^{\max}$ (see right-hand side inset of Fig.\thinspace
\ref{Fig5-Al-granular-PhaseDiagram}). When the Kondo effect is weak, it is
difficult to distinguish between $T_{\text{K}}^{\max}$ and $T_{c}$ or
$T_{c}^{onset}$. In this work we followed the evolution of $T_{c}^{onset}%
(\rho_{\text{300K}},n\mathrm{th)}$ within the region starting just before the
strong surge of Kondo behavior. Figure \thinspace
\ref{Fig5-Al-granular-PhaseDiagram} indicates that the extrapolation of this
$T_{c}^{onset}(\rho_{\text{300K}})$ agrees satisfactorily with the evolution
of $T_{\text{K}}^{\max}(\rho_{\text{300K}})$ reported for granular
films.\cite{Bachar15-Mott-granular-Al,*Bachar13-Kondo-granular-Al,Bachar14-PhD-Thesis}
We identify this range, $T_{c}^{zero}\leq T\leq T_{\text{K}}^{\max}$, of
Fig.\thinspace\ref{Fig5-Al-granular-PhaseDiagram}\ as being a superconducting
fluctuation region.

In summary, an incorporation of a chemically active oxygen in Al thin films
leads to a negative curvature resistivity, Kondo behavior, and enhancement of
$T_{c}$. The obtained $T-x$ phase diagram is shown to be similar in both the
granular and irradiated films. The driving mechanisms behind each of the
involved processes as well as the aging effects are discussed.

\begin{acknowledgments}
We gratefully acknowledge the technical assistance of K. S. F. M. Ara\'{u}jo
and the use of the facilities at LABNANO-CBPF. Partial financial support from
the Brazilian agencies CNPq and FAPERJ is also gratefully acknowledged.
\end{acknowledgments}

\bibliographystyle{apsrev4-1}
\bibliography{FermiLiquid-HeavyFermion-Kondo,intermetallic,InterplaySupMag,Localization,pnictides,SupClassic,ThinFilms}

\begin{thebibliography}{49}%
\makeatletter
\providecommand \@ifxundefined [1]{%
 \@ifx{#1\undefined}
}%
\providecommand \@ifnum [1]{%
 \ifnum #1\expandafter \@firstoftwo
 \else \expandafter \@secondoftwo
 \fi
}%
\providecommand \@ifx [1]{%
 \ifx #1\expandafter \@firstoftwo
 \else \expandafter \@secondoftwo
 \fi
}%
\providecommand \natexlab [1]{#1}%
\providecommand \enquote  [1]{``#1''}%
\providecommand \bibnamefont  [1]{#1}%
\providecommand \bibfnamefont [1]{#1}%
\providecommand \citenamefont [1]{#1}%
\providecommand \href@noop [0]{\@secondoftwo}%
\providecommand \href [0]{\begingroup \@sanitize@url \@href}%
\providecommand \@href[1]{\@@startlink{#1}\@@href}%
\providecommand \@@href[1]{\endgroup#1\@@endlink}%
\providecommand \@sanitize@url [0]{\catcode `\\12\catcode `\$12\catcode
  `\&12\catcode `\#12\catcode `\^12\catcode `\_12\catcode `\%12\relax}%
\providecommand \@@startlink[1]{}%
\providecommand \@@endlink[0]{}%
\providecommand \url  [0]{\begingroup\@sanitize@url \@url }%
\providecommand \@url [1]{\endgroup\@href {#1}{\urlprefix }}%
\providecommand \urlprefix  [0]{URL }%
\providecommand \Eprint [0]{\href }%
\providecommand \doibase [0]{http://dx.doi.org/}%
\providecommand \selectlanguage [0]{\@gobble}%
\providecommand \bibinfo  [0]{\@secondoftwo}%
\providecommand \bibfield  [0]{\@secondoftwo}%
\providecommand \translation [1]{[#1]}%
\providecommand \BibitemOpen [0]{}%
\providecommand \bibitemStop [0]{}%
\providecommand \bibitemNoStop [0]{.\EOS\space}%
\providecommand \EOS [0]{\spacefactor3000\relax}%
\providecommand \BibitemShut  [1]{\csname bibitem#1\endcsname}%
\let\auto@bib@innerbib\@empty
\bibitem [{\citenamefont {Abeles}\ \emph {et~al.}(1975)\citenamefont {Abeles},
  \citenamefont {Sheng}, \citenamefont {Coutts},\ and\ \citenamefont
  {Arie}}]{Abeles75-Granular-Films-Review}%
  \BibitemOpen
  \bibfield  {author} {\bibinfo {author} {\bibfnamefont {B.}~\bibnamefont
  {Abeles}}, \bibinfo {author} {\bibfnamefont {P.}~\bibnamefont {Sheng}},
  \bibinfo {author} {\bibfnamefont {M.}~\bibnamefont {Coutts}}, \ and\ \bibinfo
  {author} {\bibfnamefont {Y.}~\bibnamefont {Arie}},\ }\href@noop {} {\bibfield
   {journal} {\bibinfo  {journal} {Adv. Phys.}\ }\textbf {\bibinfo {volume}
  {24}},\ \bibinfo {pages} {407} (\bibinfo {year} {1975})}\BibitemShut
  {NoStop}%
\bibitem [{\citenamefont
  {Deutscher}(2008)}]{Deutscher08-Nano-Granular-SUC-Review}%
  \BibitemOpen
  \bibfield  {author} {\bibinfo {author} {\bibfnamefont {G.}~\bibnamefont
  {Deutscher}},\ }in\ \href@noop {} {\emph {\bibinfo {booktitle}
  {Superconductivity}}},\ \bibinfo {editor} {edited by\ \bibinfo {editor}
  {\bibfnamefont {K.}~\bibnamefont {Bennemann}}\ and\ \bibinfo {editor}
  {\bibfnamefont {J.}~\bibnamefont {Ketterson}}}\ (\bibinfo  {publisher}
  {Springer, Berlin},\ \bibinfo {year} {2008})\ p.\ \bibinfo {pages}
  {259}\BibitemShut {NoStop}%
\bibitem [{\citenamefont {Bachar}\ \emph {et~al.}(2015)\citenamefont {Bachar},
  \citenamefont {Lerer}, \citenamefont {Levy}, \citenamefont {Hacohen-Gourgy},
  \citenamefont {Almog}, \citenamefont {Saadaoui}, \citenamefont {Salman},
  \citenamefont {Morenzoni},\ and\ \citenamefont
  {Deutscher}}]{Bachar15-Mott-granular-Al}%
  \BibitemOpen
  \bibfield  {author} {\bibinfo {author} {\bibfnamefont {N.}~\bibnamefont
  {Bachar}}, \bibinfo {author} {\bibfnamefont {S.}~\bibnamefont {Lerer}},
  \bibinfo {author} {\bibfnamefont {A.}~\bibnamefont {Levy}}, \bibinfo {author}
  {\bibfnamefont {S.}~\bibnamefont {Hacohen-Gourgy}}, \bibinfo {author}
  {\bibfnamefont {B.}~\bibnamefont {Almog}}, \bibinfo {author} {\bibfnamefont
  {H.}~\bibnamefont {Saadaoui}}, \bibinfo {author} {\bibfnamefont
  {Z.}~\bibnamefont {Salman}}, \bibinfo {author} {\bibfnamefont
  {E.}~\bibnamefont {Morenzoni}}, \ and\ \bibinfo {author} {\bibfnamefont
  {G.}~\bibnamefont {Deutscher}},\ }\href@noop {} {\bibfield  {journal}
  {\bibinfo  {journal} {Phys. Rev. B}\ }\textbf {\bibinfo {volume} {91}},\
  \bibinfo {pages} {041123} (\bibinfo {year} {2015})}\BibitemShut {NoStop}%
\bibitem [{\citenamefont {Bachar}\ \emph {et~al.}(2013)\citenamefont {Bachar},
  \citenamefont {Lerer}, \citenamefont {Hacohen-Gourgy}, \citenamefont
  {Almog},\ and\ \citenamefont {Deutscher}}]{Bachar13-Kondo-granular-Al}%
  \BibitemOpen
  \bibfield  {author} {\bibinfo {author} {\bibfnamefont {N.}~\bibnamefont
  {Bachar}}, \bibinfo {author} {\bibfnamefont {S.}~\bibnamefont {Lerer}},
  \bibinfo {author} {\bibfnamefont {S.}~\bibnamefont {Hacohen-Gourgy}},
  \bibinfo {author} {\bibfnamefont {B.}~\bibnamefont {Almog}}, \ and\ \bibinfo
  {author} {\bibfnamefont {G.}~\bibnamefont {Deutscher}},\ }\href@noop {}
  {\bibfield  {journal} {\bibinfo  {journal} {Phys. Rev. B}\ }\textbf {\bibinfo
  {volume} {87}},\ \bibinfo {pages} {214512} (\bibinfo {year}
  {2013})}\BibitemShut {NoStop}%
\bibitem [{\citenamefont {Pracht}\ \emph {et~al.}(2016)\citenamefont {Pracht},
  \citenamefont {Bachar}, \citenamefont {Benfatto}, \citenamefont {Deutscher},
  \citenamefont {Farber}, \citenamefont {Dressel},\ and\ \citenamefont
  {Scheffler}}]{Pracht15-SUC-dome-Granular-Al}%
  \BibitemOpen
  \bibfield  {author} {\bibinfo {author} {\bibfnamefont {U.~S.}\ \bibnamefont
  {Pracht}}, \bibinfo {author} {\bibfnamefont {N.}~\bibnamefont {Bachar}},
  \bibinfo {author} {\bibfnamefont {L.}~\bibnamefont {Benfatto}}, \bibinfo
  {author} {\bibfnamefont {G.}~\bibnamefont {Deutscher}}, \bibinfo {author}
  {\bibfnamefont {E.}~\bibnamefont {Farber}}, \bibinfo {author} {\bibfnamefont
  {M.}~\bibnamefont {Dressel}}, \ and\ \bibinfo {author} {\bibfnamefont
  {M.}~\bibnamefont {Scheffler}},\ }\href {\doibase 10.1103/PhysRevB.93.100503}
  {\bibfield  {journal} {\bibinfo  {journal} {Phys. Rev. B}\ }\textbf {\bibinfo
  {volume} {93}},\ \bibinfo {pages} {100503} (\bibinfo {year}
  {2016})}\BibitemShut {NoStop}%
\bibitem [{\citenamefont {Bachar}(2014)}]{Bachar14-PhD-Thesis}%
  \BibitemOpen
  \bibfield  {author} {\bibinfo {author} {\bibfnamefont {N.}~\bibnamefont
  {Bachar}},\ }\emph {\bibinfo {title} {Spin-flip scattering in superconducting
  granular aluminum films}},\ \href@noop {} {Ph.D. thesis},\ \bibinfo  {school}
  {Tel-Aviv University, Tel-Aviv, Israel} (\bibinfo {year} {2014}),\ \bibinfo
  {note} {unpublished}\BibitemShut {NoStop}%
\bibitem [{\citenamefont {Abeles}\ \emph {et~al.}(1967)\citenamefont {Abeles},
  \citenamefont {Cohen},\ and\ \citenamefont
  {Stowell}}]{Abeles67-Hc-granular-SUCs}%
  \BibitemOpen
  \bibfield  {author} {\bibinfo {author} {\bibfnamefont {B.}~\bibnamefont
  {Abeles}}, \bibinfo {author} {\bibfnamefont {R.~W.}\ \bibnamefont {Cohen}}, \
  and\ \bibinfo {author} {\bibfnamefont {W.~R.}\ \bibnamefont {Stowell}},\
  }\href@noop {} {\bibfield  {journal} {\bibinfo  {journal} {Phys. Rev. Lett.}\
  }\textbf {\bibinfo {volume} {18}},\ \bibinfo {pages} {902} (\bibinfo {year}
  {1967})}\BibitemShut {NoStop}%
\bibitem [{\citenamefont {Lamoise}\ \emph
  {et~al.}(1975{\natexlab{a}})\citenamefont {Lamoise}, \citenamefont
  {Chaumont}, \citenamefont {Meunier},\ and\ \citenamefont
  {Bernas}}]{Lamoise75-Irradiation-Al-SUC}%
  \BibitemOpen
  \bibfield  {author} {\bibinfo {author} {\bibfnamefont {A.}~\bibnamefont
  {Lamoise}}, \bibinfo {author} {\bibfnamefont {J.}~\bibnamefont {Chaumont}},
  \bibinfo {author} {\bibfnamefont {F.}~\bibnamefont {Meunier}}, \ and\
  \bibinfo {author} {\bibfnamefont {H.}~\bibnamefont {Bernas}},\ }\href@noop {}
  {\bibfield  {journal} {\bibinfo  {journal} {J. Phys., Lett.}\ }\textbf
  {\bibinfo {volume} {36}},\ \bibinfo {pages} {271} (\bibinfo {year}
  {1975}{\natexlab{a}})}\BibitemShut {NoStop}%
\bibitem [{\citenamefont {Lamoise}\ \emph {et~al.}(1976)\citenamefont
  {Lamoise}, \citenamefont {Chaumont}, \citenamefont {Lalu}, \citenamefont
  {Meunier},\ and\ \citenamefont {Bernas}}]{Lamoise76-Enhanced-SUC-Al-Alloys}%
  \BibitemOpen
  \bibfield  {author} {\bibinfo {author} {\bibfnamefont {A.}~\bibnamefont
  {Lamoise}}, \bibinfo {author} {\bibfnamefont {J.}~\bibnamefont {Chaumont}},
  \bibinfo {author} {\bibfnamefont {F.}~\bibnamefont {Lalu}}, \bibinfo {author}
  {\bibfnamefont {F.}~\bibnamefont {Meunier}}, \ and\ \bibinfo {author}
  {\bibfnamefont {H.}~\bibnamefont {Bernas}},\ }\href@noop {} {\bibfield
  {journal} {\bibinfo  {journal} {J. Phys., Lett.}\ }\textbf {\bibinfo {volume}
  {37}},\ \bibinfo {pages} {287} (\bibinfo {year} {1976})}\BibitemShut
  {NoStop}%
\bibitem [{\citenamefont {Lamoise}\ \emph
  {et~al.}(1975{\natexlab{b}})\citenamefont {Lamoise}, \citenamefont
  {Chaumont}, \citenamefont {Meunier},\ and\ \citenamefont
  {Bernas}}]{Lamoise75-Irradiation-Al-Resistivity-annealing}%
  \BibitemOpen
  \bibfield  {author} {\bibinfo {author} {\bibfnamefont {A.}~\bibnamefont
  {Lamoise}}, \bibinfo {author} {\bibfnamefont {J.}~\bibnamefont {Chaumont}},
  \bibinfo {author} {\bibfnamefont {F.}~\bibnamefont {Meunier}}, \ and\
  \bibinfo {author} {\bibfnamefont {H.}~\bibnamefont {Bernas}},\ }\href@noop {}
  {\bibfield  {journal} {\bibinfo  {journal} {J. Phys., Lett.}\ }\textbf
  {\bibinfo {volume} {36}},\ \bibinfo {pages} {305} (\bibinfo {year}
  {1975}{\natexlab{b}})}\BibitemShut {NoStop}%
\bibitem [{\citenamefont {Meunier}\ \emph {et~al.}(1977)\citenamefont
  {Meunier}, \citenamefont {Pfeuty}, \citenamefont {Lamoise}, \citenamefont
  {Chaumont}, \citenamefont {Bernas},\ and\ \citenamefont
  {Cohen}}]{Meunier77-Irradiation-Al-Percolation-SUC}%
  \BibitemOpen
  \bibfield  {author} {\bibinfo {author} {\bibfnamefont {F.}~\bibnamefont
  {Meunier}}, \bibinfo {author} {\bibfnamefont {P.}~\bibnamefont {Pfeuty}},
  \bibinfo {author} {\bibfnamefont {A.}~\bibnamefont {Lamoise}}, \bibinfo
  {author} {\bibfnamefont {J.}~\bibnamefont {Chaumont}}, \bibinfo {author}
  {\bibfnamefont {H.}~\bibnamefont {Bernas}}, \ and\ \bibinfo {author}
  {\bibfnamefont {C.}~\bibnamefont {Cohen}},\ }\href@noop {} {\bibfield
  {journal} {\bibinfo  {journal} {J. Phys., Lett.}\ }\textbf {\bibinfo {volume}
  {38}},\ \bibinfo {pages} {435} (\bibinfo {year} {1977})}\BibitemShut
  {NoStop}%
\bibitem [{\citenamefont {Deutscher}\ \emph
  {et~al.}(1973{\natexlab{a}})\citenamefont {Deutscher}, \citenamefont
  {Gershenson}, \citenamefont {Grunbaum},\ and\ \citenamefont
  {Imry}}]{Deutscher73-Granular-SUC-Films}%
  \BibitemOpen
  \bibfield  {author} {\bibinfo {author} {\bibfnamefont {G.}~\bibnamefont
  {Deutscher}}, \bibinfo {author} {\bibfnamefont {M.}~\bibnamefont
  {Gershenson}}, \bibinfo {author} {\bibfnamefont {E.}~\bibnamefont
  {Grunbaum}}, \ and\ \bibinfo {author} {\bibfnamefont {Y.}~\bibnamefont
  {Imry}},\ }\href@noop {} {\bibfield  {journal} {\bibinfo  {journal} {J. Vac.
  Sci. Technol.}\ }\textbf {\bibinfo {volume} {10}},\ \bibinfo {pages} {697}
  (\bibinfo {year} {1973}{\natexlab{a}})}\BibitemShut {NoStop}%
\bibitem [{\citenamefont {Deutscher}\ \emph
  {et~al.}(1973{\natexlab{b}})\citenamefont {Deutscher}, \citenamefont
  {Fenichel}, \citenamefont {Gershenson}, \citenamefont {Gr{\"u}nbaum},\ and\
  \citenamefont {Ovadyahu}}]{Deutscher73-Granular-SUC-zeroDimension}%
  \BibitemOpen
  \bibfield  {author} {\bibinfo {author} {\bibfnamefont {G.}~\bibnamefont
  {Deutscher}}, \bibinfo {author} {\bibfnamefont {H.}~\bibnamefont {Fenichel}},
  \bibinfo {author} {\bibfnamefont {M.}~\bibnamefont {Gershenson}}, \bibinfo
  {author} {\bibfnamefont {E.}~\bibnamefont {Gr{\"u}nbaum}}, \ and\ \bibinfo
  {author} {\bibfnamefont {Z.}~\bibnamefont {Ovadyahu}},\ }\href@noop {}
  {\bibfield  {journal} {\bibinfo  {journal} {J. Low Temp. Phys.}\ }\textbf
  {\bibinfo {volume} {10}},\ \bibinfo {pages} {231} (\bibinfo {year}
  {1973}{\natexlab{b}})}\BibitemShut {NoStop}%
\bibitem [{\citenamefont {Isebeck}\ \emph {et~al.}(1966)\citenamefont
  {Isebeck}, \citenamefont {Muller}, \citenamefont {Schilling},\ and\
  \citenamefont {Wenzl}}]{Isebeck66-Reconvery-in-Al-NeutronIrradiation}%
  \BibitemOpen
  \bibfield  {author} {\bibinfo {author} {\bibfnamefont {K.}~\bibnamefont
  {Isebeck}}, \bibinfo {author} {\bibfnamefont {R.}~\bibnamefont {Muller}},
  \bibinfo {author} {\bibfnamefont {W.}~\bibnamefont {Schilling}}, \ and\
  \bibinfo {author} {\bibfnamefont {H.}~\bibnamefont {Wenzl}},\ }\href@noop {}
  {\bibfield  {journal} {\bibinfo  {journal} {Phys. Stat. Sol. 18}\ }\textbf
  {\bibinfo {volume} {18}},\ \bibinfo {pages} {427} (\bibinfo {year}
  {1966})}\BibitemShut {NoStop}%
\bibitem [{\citenamefont {Sosin}\ and\ \citenamefont
  {Rachal}(1963)}]{Sosin63-Aluminum-Recovery-AfterIrradiation}%
  \BibitemOpen
  \bibfield  {author} {\bibinfo {author} {\bibfnamefont {A.}~\bibnamefont
  {Sosin}}\ and\ \bibinfo {author} {\bibfnamefont {L.~H.}\ \bibnamefont
  {Rachal}},\ }\href {\doibase 10.1103/PhysRev.130.2238} {\bibfield  {journal}
  {\bibinfo  {journal} {Phys. Rev.}\ }\textbf {\bibinfo {volume} {130}},\
  \bibinfo {pages} {2238} (\bibinfo {year} {1963})}\BibitemShut {NoStop}%
\bibitem [{\citenamefont {DeSorbo}\ and\ \citenamefont
  {Turnbull}(1959)}]{DeSorbo59-Kinetics-Vacancy-Al}%
  \BibitemOpen
  \bibfield  {author} {\bibinfo {author} {\bibfnamefont {W.}~\bibnamefont
  {DeSorbo}}\ and\ \bibinfo {author} {\bibfnamefont {D.}~\bibnamefont
  {Turnbull}},\ }\href@noop {} {\bibfield  {journal} {\bibinfo  {journal}
  {Phys. Rev.}\ }\textbf {\bibinfo {volume} {115}},\ \bibinfo {pages} {560}
  (\bibinfo {year} {1959})}\BibitemShut {NoStop}%
\bibitem [{\citenamefont {Khellaf}\ \emph {et~al.}(2002)\citenamefont
  {Khellaf}, \citenamefont {Seeger},\ and\ \citenamefont
  {Emrick}}]{Khellaf02-Quench-Aluminum}%
  \BibitemOpen
  \bibfield  {author} {\bibinfo {author} {\bibfnamefont {A.}~\bibnamefont
  {Khellaf}}, \bibinfo {author} {\bibfnamefont {A.}~\bibnamefont {Seeger}}, \
  and\ \bibinfo {author} {\bibfnamefont {R.~M.}\ \bibnamefont {Emrick}},\
  }\href@noop {} {\bibfield  {journal} {\bibinfo  {journal} {Mater. Trans.}\
  }\textbf {\bibinfo {volume} {43}},\ \bibinfo {pages} {186} (\bibinfo {year}
  {2002})}\BibitemShut {NoStop}%
\bibitem [{\citenamefont {Mello}\ \emph {et~al.}(2016)\citenamefont {Mello},
  \citenamefont {Codeco}, \citenamefont {Magnani},\ and\ \citenamefont
  {Sant'Anna}}]{Mello16-Accelerator-SIMS}%
  \BibitemOpen
  \bibfield  {author} {\bibinfo {author} {\bibfnamefont {S.~L.~A.}\
  \bibnamefont {Mello}}, \bibinfo {author} {\bibfnamefont {C.~F.~S.}\
  \bibnamefont {Codeco}}, \bibinfo {author} {\bibfnamefont {B.~F.}\
  \bibnamefont {Magnani}}, \ and\ \bibinfo {author} {\bibfnamefont {M.~M.}\
  \bibnamefont {Sant'Anna}},\ }\href@noop {} {\bibfield  {journal} {\bibinfo
  {journal} {Rev. Sci. Instrum.}\ }\textbf {\bibinfo {volume} {87}},\ \bibinfo
  {pages} {063305} (\bibinfo {year} {2016})}\BibitemShut {NoStop}%
\bibitem [{\citenamefont {Bandyopadhyay}\ \emph {et~al.}(1982)\citenamefont
  {Bandyopadhyay}, \citenamefont {Lindenfeld}, \citenamefont {McLean},\ and\
  \citenamefont {Sin}}]{Bandyopadhyay82-HallEffect-granular-Al}%
  \BibitemOpen
  \bibfield  {author} {\bibinfo {author} {\bibfnamefont {B.}~\bibnamefont
  {Bandyopadhyay}}, \bibinfo {author} {\bibfnamefont {P.}~\bibnamefont
  {Lindenfeld}}, \bibinfo {author} {\bibfnamefont {W.~L.}\ \bibnamefont
  {McLean}}, \ and\ \bibinfo {author} {\bibfnamefont {H.~K.}\ \bibnamefont
  {Sin}},\ }\href@noop {} {\bibfield  {journal} {\bibinfo  {journal} {Phys.
  Rev. B}\ }\textbf {\bibinfo {volume} {26}},\ \bibinfo {pages} {3476}
  (\bibinfo {year} {1982})}\BibitemShut {NoStop}%
\bibitem [{\citenamefont {Ziegler}\ \emph {et~al.}(2010)\citenamefont
  {Ziegler}, \citenamefont {Ziegler},\ and\ \citenamefont
  {Biersack}}]{Ziegler10-SRIM}%
  \BibitemOpen
  \bibfield  {author} {\bibinfo {author} {\bibfnamefont {J.~F.}\ \bibnamefont
  {Ziegler}}, \bibinfo {author} {\bibfnamefont {M.}~\bibnamefont {Ziegler}}, \
  and\ \bibinfo {author} {\bibfnamefont {J.}~\bibnamefont {Biersack}},\
  }\href@noop {} {\bibfield  {journal} {\bibinfo  {journal} {Nucl. Instr. Meth.
  B}\ }\textbf {\bibinfo {volume} {268}},\ \bibinfo {pages} {1818} (\bibinfo
  {year} {2010})}\BibitemShut {NoStop}%
\bibitem [{\citenamefont {Boggio}\ and\ \citenamefont
  {Plumb}(1966)}]{Boggio66-Metal-ThinFilms-Theory}%
  \BibitemOpen
  \bibfield  {author} {\bibinfo {author} {\bibfnamefont {J.~E.}\ \bibnamefont
  {Boggio}}\ and\ \bibinfo {author} {\bibfnamefont {R.~C.}\ \bibnamefont
  {Plumb}},\ }\href@noop {} {\bibfield  {journal} {\bibinfo  {journal} {J.
  Chem. Phys.}\ }\textbf {\bibinfo {volume} {44}},\ \bibinfo {pages} {1081}
  (\bibinfo {year} {1966})}\BibitemShut {NoStop}%
\bibitem [{\citenamefont {Atkinson}(1985)}]{Atkinson85-Growth-Oxides-Films}%
  \BibitemOpen
  \bibfield  {author} {\bibinfo {author} {\bibfnamefont {A.}~\bibnamefont
  {Atkinson}},\ }\href@noop {} {\bibfield  {journal} {\bibinfo  {journal} {Rev.
  Mod. Phys.}\ }\textbf {\bibinfo {volume} {57}},\ \bibinfo {pages} {437}
  (\bibinfo {year} {1985})}\BibitemShut {NoStop}%
\bibitem [{\citenamefont {Dorey}\ and\ \citenamefont
  {Knight}(1969)}]{Dorey69-Resistivity-vs-time}%
  \BibitemOpen
  \bibfield  {author} {\bibinfo {author} {\bibfnamefont {A.}~\bibnamefont
  {Dorey}}\ and\ \bibinfo {author} {\bibfnamefont {J.}~\bibnamefont {Knight}},\
  }\href {\doibase http://dx.doi.org/10.1016/0040-6090(69)90093-5} {\bibfield
  {journal} {\bibinfo  {journal} {Thin Solid Films}\ }\textbf {\bibinfo
  {volume} {4}},\ \bibinfo {pages} {445 } (\bibinfo {year} {1969})}\BibitemShut
  {NoStop}%
\bibitem [{\citenamefont {Day}\ \emph {et~al.}(1995)\citenamefont {Day},
  \citenamefont {Delfino}, \citenamefont {Fair},\ and\ \citenamefont
  {Tsai}}]{Day95-resistivity-GrainSize-Ti-Films}%
  \BibitemOpen
  \bibfield  {author} {\bibinfo {author} {\bibfnamefont {M.~E.}\ \bibnamefont
  {Day}}, \bibinfo {author} {\bibfnamefont {M.}~\bibnamefont {Delfino}},
  \bibinfo {author} {\bibfnamefont {J.~A.}\ \bibnamefont {Fair}}, \ and\
  \bibinfo {author} {\bibfnamefont {W.}~\bibnamefont {Tsai}},\ }\href@noop {}
  {\bibfield  {journal} {\bibinfo  {journal} {Thin Solid Films}\ }\textbf
  {\bibinfo {volume} {254}},\ \bibinfo {pages} {285} (\bibinfo {year}
  {1995})}\BibitemShut {NoStop}%
\bibitem [{\citenamefont {Cabrera}\ and\ \citenamefont
  {Mott}(1949)}]{Cabrera49-oxidation-ThinFims}%
  \BibitemOpen
  \bibfield  {author} {\bibinfo {author} {\bibfnamefont {N.}~\bibnamefont
  {Cabrera}}\ and\ \bibinfo {author} {\bibfnamefont {N.~F.}\ \bibnamefont
  {Mott}},\ }\href@noop {} {\bibfield  {journal} {\bibinfo  {journal} {Rep.
  Prog. Phys.}\ }\textbf {\bibinfo {volume} {12}},\ \bibinfo {pages} {163}
  (\bibinfo {year} {1949})}\BibitemShut {NoStop}%
\bibitem [{\citenamefont {Gamble}\ \emph {et~al.}(1964)\citenamefont {Gamble},
  \citenamefont {Bartram}, \citenamefont {Young}, \citenamefont {Gilliam},\
  and\ \citenamefont {Levy}}]{Gamble64-EPR-Al2P3}%
  \BibitemOpen
  \bibfield  {author} {\bibinfo {author} {\bibfnamefont {F.~T.}\ \bibnamefont
  {Gamble}}, \bibinfo {author} {\bibfnamefont {R.~H.}\ \bibnamefont {Bartram}},
  \bibinfo {author} {\bibfnamefont {C.~G.}\ \bibnamefont {Young}}, \bibinfo
  {author} {\bibfnamefont {O.~R.}\ \bibnamefont {Gilliam}}, \ and\ \bibinfo
  {author} {\bibfnamefont {P.~W.}\ \bibnamefont {Levy}},\ }\href@noop {}
  {\bibfield  {journal} {\bibinfo  {journal} {Phys. Rev.}\ }\textbf {\bibinfo
  {volume} {134}},\ \bibinfo {pages} {A589} (\bibinfo {year}
  {1964})}\BibitemShut {NoStop}%
\bibitem [{\citenamefont {Lee}\ \emph {et~al.}(1976)\citenamefont {Lee},
  \citenamefont {Holmberg},\ and\ \citenamefont
  {Crawford}}]{Lee76-holes-Al2O3}%
  \BibitemOpen
  \bibfield  {author} {\bibinfo {author} {\bibfnamefont {K.}~\bibnamefont
  {Lee}}, \bibinfo {author} {\bibfnamefont {G.}~\bibnamefont {Holmberg}}, \
  and\ \bibinfo {author} {\bibfnamefont {J.}~\bibnamefont {Crawford}},\
  }\href@noop {} {\bibfield  {journal} {\bibinfo  {journal} {Solid State
  Commun.}\ }\textbf {\bibinfo {volume} {20}},\ \bibinfo {pages} {183}
  (\bibinfo {year} {1976})}\BibitemShut {NoStop}%
\bibitem [{\citenamefont {Chui}\ \emph
  {et~al.}(1981{\natexlab{a}})\citenamefont {Chui}, \citenamefont {Lindenfeld},
  \citenamefont {McLean},\ and\ \citenamefont
  {Mui}}]{Chui81-Nano-Al-MagRes-Localizatio}%
  \BibitemOpen
  \bibfield  {author} {\bibinfo {author} {\bibfnamefont {T.}~\bibnamefont
  {Chui}}, \bibinfo {author} {\bibfnamefont {P.}~\bibnamefont {Lindenfeld}},
  \bibinfo {author} {\bibfnamefont {W.~L.}\ \bibnamefont {McLean}}, \ and\
  \bibinfo {author} {\bibfnamefont {K.}~\bibnamefont {Mui}},\ }\href@noop {}
  {\bibfield  {journal} {\bibinfo  {journal} {Phys. Rev. Lett.}\ }\textbf
  {\bibinfo {volume} {47}},\ \bibinfo {pages} {1617} (\bibinfo {year}
  {1981}{\natexlab{a}})}\BibitemShut {NoStop}%
\bibitem [{\citenamefont {Chui}\ \emph
  {et~al.}(1981{\natexlab{b}})\citenamefont {Chui}, \citenamefont {Lindenfeld},
  \citenamefont {McLean},\ and\ \citenamefont
  {Mui}}]{Chui81-Tc-Hc-Granular-Al-SUC}%
  \BibitemOpen
  \bibfield  {author} {\bibinfo {author} {\bibfnamefont {T.}~\bibnamefont
  {Chui}}, \bibinfo {author} {\bibfnamefont {P.}~\bibnamefont {Lindenfeld}},
  \bibinfo {author} {\bibfnamefont {W.~L.}\ \bibnamefont {McLean}}, \ and\
  \bibinfo {author} {\bibfnamefont {K.}~\bibnamefont {Mui}},\ }\href@noop {}
  {\bibfield  {journal} {\bibinfo  {journal} {Phys. Rev. B}\ }\textbf {\bibinfo
  {volume} {24}},\ \bibinfo {pages} {6728} (\bibinfo {year}
  {1981}{\natexlab{b}})}\BibitemShut {NoStop}%
\bibitem [{\citenamefont {Mui}\ \emph {et~al.}(1984)\citenamefont {Mui},
  \citenamefont {Lindenfeld},\ and\ \citenamefont
  {McLean}}]{Mui84-Granular-Al-MagRes}%
  \BibitemOpen
  \bibfield  {author} {\bibinfo {author} {\bibfnamefont {K.~C.}\ \bibnamefont
  {Mui}}, \bibinfo {author} {\bibfnamefont {P.}~\bibnamefont {Lindenfeld}}, \
  and\ \bibinfo {author} {\bibfnamefont {W.~L.}\ \bibnamefont {McLean}},\
  }\href@noop {} {\bibfield  {journal} {\bibinfo  {journal} {Phys. Rev. B}\
  }\textbf {\bibinfo {volume} {30}},\ \bibinfo {pages} {2951} (\bibinfo {year}
  {1984})}\BibitemShut {NoStop}%
\bibitem [{\citenamefont {Desai}\ \emph {et~al.}(1984)\citenamefont {Desai},
  \citenamefont {James},\ and\ \citenamefont {Ho}}]{Desai84-Al-resistivity}%
  \BibitemOpen
  \bibfield  {author} {\bibinfo {author} {\bibfnamefont {P.~D.}\ \bibnamefont
  {Desai}}, \bibinfo {author} {\bibfnamefont {H.~M.}\ \bibnamefont {James}}, \
  and\ \bibinfo {author} {\bibfnamefont {C.~Y.}\ \bibnamefont {Ho}},\
  }\href@noop {} {\bibfield  {journal} {\bibinfo  {journal} {J. Phys. Chem.
  Ref. Data Rep.}\ }\textbf {\bibinfo {volume} {13}},\ \bibinfo {pages} {1131}
  (\bibinfo {year} {1984})}\BibitemShut {NoStop}%
\bibitem [{\citenamefont {Lee}\ \emph {et~al.}(2014)\citenamefont {Lee},
  \citenamefont {DuBois},\ and\ \citenamefont
  {Lordi}}]{Lee14-Mag-Al2O3-Al-interface}%
  \BibitemOpen
  \bibfield  {author} {\bibinfo {author} {\bibfnamefont {D.}~\bibnamefont
  {Lee}}, \bibinfo {author} {\bibfnamefont {J.~L.}\ \bibnamefont {DuBois}}, \
  and\ \bibinfo {author} {\bibfnamefont {V.}~\bibnamefont {Lordi}},\
  }\href@noop {} {\bibfield  {journal} {\bibinfo  {journal} {Phys. Rev. Lett.}\
  }\textbf {\bibinfo {volume} {112}},\ \bibinfo {pages} {017001} (\bibinfo
  {year} {2014})}\BibitemShut {NoStop}%
\bibitem [{\citenamefont {Anton}\ \emph {et~al.}(2013)\citenamefont {Anton},
  \citenamefont {Birenbaum}, \citenamefont {O'Kelley}, \citenamefont
  {Bolkhovsky}, \citenamefont {Braje}, \citenamefont {Fitch}, \citenamefont
  {Neeley}, \citenamefont {Hilton}, \citenamefont {Cho}, \citenamefont {Irwin},
  \citenamefont {Wellstood}, \citenamefont {Oliver}, \citenamefont {Shnirman},\
  and\ \citenamefont {Clarke}}]{Anton13-SQUID-Noise}%
  \BibitemOpen
  \bibfield  {author} {\bibinfo {author} {\bibfnamefont {S.~M.}\ \bibnamefont
  {Anton}}, \bibinfo {author} {\bibfnamefont {J.~S.}\ \bibnamefont
  {Birenbaum}}, \bibinfo {author} {\bibfnamefont {S.~R.}\ \bibnamefont
  {O'Kelley}}, \bibinfo {author} {\bibfnamefont {V.}~\bibnamefont
  {Bolkhovsky}}, \bibinfo {author} {\bibfnamefont {D.~A.}\ \bibnamefont
  {Braje}}, \bibinfo {author} {\bibfnamefont {G.}~\bibnamefont {Fitch}},
  \bibinfo {author} {\bibfnamefont {M.}~\bibnamefont {Neeley}}, \bibinfo
  {author} {\bibfnamefont {G.~C.}\ \bibnamefont {Hilton}}, \bibinfo {author}
  {\bibfnamefont {H.-M.}\ \bibnamefont {Cho}}, \bibinfo {author} {\bibfnamefont
  {K.~D.}\ \bibnamefont {Irwin}}, \bibinfo {author} {\bibfnamefont {F.~C.}\
  \bibnamefont {Wellstood}}, \bibinfo {author} {\bibfnamefont {W.~D.}\
  \bibnamefont {Oliver}}, \bibinfo {author} {\bibfnamefont {A.}~\bibnamefont
  {Shnirman}}, \ and\ \bibinfo {author} {\bibfnamefont {J.}~\bibnamefont
  {Clarke}},\ }\href@noop {} {\bibfield  {journal} {\bibinfo  {journal} {Phys.
  Rev. Lett.}\ }\textbf {\bibinfo {volume} {110}},\ \bibinfo {pages} {147002}
  (\bibinfo {year} {2013})}\BibitemShut {NoStop}%
\bibitem [{\citenamefont {Bluhm}\ \emph {et~al.}(2009)\citenamefont {Bluhm},
  \citenamefont {Bert}, \citenamefont {Koshnick}, \citenamefont {Huber},\ and\
  \citenamefont {Moler}}]{Bluhm09-MagSuc-Al-Film}%
  \BibitemOpen
  \bibfield  {author} {\bibinfo {author} {\bibfnamefont {H.}~\bibnamefont
  {Bluhm}}, \bibinfo {author} {\bibfnamefont {J.~A.}\ \bibnamefont {Bert}},
  \bibinfo {author} {\bibfnamefont {N.~C.}\ \bibnamefont {Koshnick}}, \bibinfo
  {author} {\bibfnamefont {M.~E.}\ \bibnamefont {Huber}}, \ and\ \bibinfo
  {author} {\bibfnamefont {K.~A.}\ \bibnamefont {Moler}},\ }\href@noop {}
  {\bibfield  {journal} {\bibinfo  {journal} {Phys. Rev. Lett.}\ }\textbf
  {\bibinfo {volume} {103}},\ \bibinfo {pages} {026805} (\bibinfo {year}
  {2009})}\BibitemShut {NoStop}%
\bibitem [{\citenamefont {Sendelbach}\ \emph {et~al.}(2008)\citenamefont
  {Sendelbach}, \citenamefont {Hover}, \citenamefont {Kittel}, \citenamefont
  {M\"uck}, \citenamefont {Martinis},\ and\ \citenamefont
  {McDermott}}]{Sendelbach08-Maggnetism-in-SQUID}%
  \BibitemOpen
  \bibfield  {author} {\bibinfo {author} {\bibfnamefont {S.}~\bibnamefont
  {Sendelbach}}, \bibinfo {author} {\bibfnamefont {D.}~\bibnamefont {Hover}},
  \bibinfo {author} {\bibfnamefont {A.}~\bibnamefont {Kittel}}, \bibinfo
  {author} {\bibfnamefont {M.}~\bibnamefont {M\"uck}}, \bibinfo {author}
  {\bibfnamefont {J.~M.}\ \bibnamefont {Martinis}}, \ and\ \bibinfo {author}
  {\bibfnamefont {R.}~\bibnamefont {McDermott}},\ }\href@noop {} {\bibfield
  {journal} {\bibinfo  {journal} {Phys. Rev. Lett.}\ }\textbf {\bibinfo
  {volume} {100}},\ \bibinfo {pages} {227006} (\bibinfo {year}
  {2008})}\BibitemShut {NoStop}%
\bibitem [{\citenamefont {Linker}\ and\ \citenamefont
  {Meyer}(1976)}]{Linker76-Tc-Mo-Implantation}%
  \BibitemOpen
  \bibfield  {author} {\bibinfo {author} {\bibfnamefont {G.}~\bibnamefont
  {Linker}}\ and\ \bibinfo {author} {\bibfnamefont {O.}~\bibnamefont {Meyer}},\
  }\href@noop {} {\bibfield  {journal} {\bibinfo  {journal} {Solid State
  Commun.}\ }\textbf {\bibinfo {volume} {20}},\ \bibinfo {pages} {695}
  (\bibinfo {year} {1976})}\BibitemShut {NoStop}%
\bibitem [{\citenamefont
  {Linker}(1982)}]{Linker82-Implantation-Mo-Tc-Enhancement}%
  \BibitemOpen
  \bibfield  {author} {\bibinfo {author} {\bibfnamefont {G.}~\bibnamefont
  {Linker}},\ }in\ \href {\doibase
  http://dx.doi.org/10.1016/B978-0-08-027625-0.50035-6} {\emph {\bibinfo
  {booktitle} {Ion Implantation Into Metals}}},\ \bibinfo {editor} {edited by\
  \bibinfo {editor} {\bibfnamefont {V.~A.~G.}\ \bibnamefont {Procter}}}\
  (\bibinfo  {publisher} {Pergamon, New York},\ \bibinfo {year} {1982})\ pp.\
  \bibinfo {pages} {284 -- 292}\BibitemShut {NoStop}%
\bibitem [{\citenamefont {Beloborodov}\ \emph {et~al.}(2007)\citenamefont
  {Beloborodov}, \citenamefont {Lopatin}, \citenamefont {Vinokur},\ and\
  \citenamefont {Efetov}}]{Beloborodov07-Transport-Granular-Review}%
  \BibitemOpen
  \bibfield  {author} {\bibinfo {author} {\bibfnamefont {I.~S.}\ \bibnamefont
  {Beloborodov}}, \bibinfo {author} {\bibfnamefont {A.~V.}\ \bibnamefont
  {Lopatin}}, \bibinfo {author} {\bibfnamefont {V.~M.}\ \bibnamefont
  {Vinokur}}, \ and\ \bibinfo {author} {\bibfnamefont {K.~B.}\ \bibnamefont
  {Efetov}},\ }\href@noop {} {\bibfield  {journal} {\bibinfo  {journal} {Rev.
  Mod. Phys.}\ }\textbf {\bibinfo {volume} {79}},\ \bibinfo {pages} {469}
  (\bibinfo {year} {2007})}\BibitemShut {NoStop}%
\bibitem [{\citenamefont {Garland}\ \emph {et~al.}(1968)\citenamefont
  {Garland}, \citenamefont {Bennemann},\ and\ \citenamefont
  {Mueller}}]{Garland68-Disorder-enhanced-Tc-Al}%
  \BibitemOpen
  \bibfield  {author} {\bibinfo {author} {\bibfnamefont {J.~W.}\ \bibnamefont
  {Garland}}, \bibinfo {author} {\bibfnamefont {K.~H.}\ \bibnamefont
  {Bennemann}}, \ and\ \bibinfo {author} {\bibfnamefont {F.~M.}\ \bibnamefont
  {Mueller}},\ }\href@noop {} {\bibfield  {journal} {\bibinfo  {journal} {Phys.
  Rev. Lett.}\ }\textbf {\bibinfo {volume} {21}},\ \bibinfo {pages} {1315}
  (\bibinfo {year} {1968})}\BibitemShut {NoStop}%
\bibitem [{\citenamefont {Klein}\ and\ \citenamefont
  {Leger}(1968)}]{Klein68-Phonon-Granular-Al}%
  \BibitemOpen
  \bibfield  {author} {\bibinfo {author} {\bibfnamefont {J.}~\bibnamefont
  {Klein}}\ and\ \bibinfo {author} {\bibfnamefont {A.}~\bibnamefont {Leger}},\
  }\href {\doibase http://dx.doi.org/10.1016/0375-9601(68)90431-3} {\bibfield
  {journal} {\bibinfo  {journal} {Phys. Lett. A}\ }\textbf {\bibinfo {volume}
  {28}},\ \bibinfo {pages} {134 } (\bibinfo {year} {1968})}\BibitemShut
  {NoStop}%
\bibitem [{\citenamefont {Leger}\ and\ \citenamefont
  {Klein}(1969)}]{LEGER69-Enhanced-SUC}%
  \BibitemOpen
  \bibfield  {author} {\bibinfo {author} {\bibfnamefont {A.}~\bibnamefont
  {Leger}}\ and\ \bibinfo {author} {\bibfnamefont {J.}~\bibnamefont {Klein}},\
  }\href {\doibase http://dx.doi.org/10.1016/0375-9601(69)90601-X} {\bibfield
  {journal} {\bibinfo  {journal} {Phys. Lett. A}\ }\textbf {\bibinfo {volume}
  {28}},\ \bibinfo {pages} {751 } (\bibinfo {year} {1969})}\BibitemShut
  {NoStop}%
\bibitem [{\citenamefont {Allender}\ \emph {et~al.}(1973)\citenamefont
  {Allender}, \citenamefont {Bray},\ and\ \citenamefont
  {Bardeen}}]{Allender73-Excitonic-SUC}%
  \BibitemOpen
  \bibfield  {author} {\bibinfo {author} {\bibfnamefont {D.}~\bibnamefont
  {Allender}}, \bibinfo {author} {\bibfnamefont {J.}~\bibnamefont {Bray}}, \
  and\ \bibinfo {author} {\bibfnamefont {J.}~\bibnamefont {Bardeen}},\ }\href
  {\doibase 10.1103/PhysRevB.7.1020} {\bibfield  {journal} {\bibinfo  {journal}
  {Phys. Rev. B}\ }\textbf {\bibinfo {volume} {7}},\ \bibinfo {pages} {1020}
  (\bibinfo {year} {1973})}\BibitemShut {NoStop}%
\bibitem [{\citenamefont {Kresin}\ and\ \citenamefont
  {Ovchinnikov}(2008)}]{Kresin08-Tc-enhanced0Nanoclusters}%
  \BibitemOpen
  \bibfield  {author} {\bibinfo {author} {\bibfnamefont {V.~Z.}\ \bibnamefont
  {Kresin}}\ and\ \bibinfo {author} {\bibfnamefont {Y.~N.}\ \bibnamefont
  {Ovchinnikov}},\ }\href@noop {} {\bibfield  {journal} {\bibinfo  {journal}
  {Physics-Uspekhi}\ }\textbf {\bibinfo {volume} {51}},\ \bibinfo {pages} {427}
  (\bibinfo {year} {2008})}\BibitemShut {NoStop}%
\bibitem [{\citenamefont {Bose}\ \emph {et~al.}(2010)\citenamefont {Bose},
  \citenamefont {García-García}, \citenamefont {Ugeda}, \citenamefont {Urbina},
  \citenamefont {Michaelis}, \citenamefont {Brihuega},\ and\ \citenamefont
  {Kern}}]{Bose10-SUC-Sn-Nanocluster}%
  \BibitemOpen
  \bibfield  {author} {\bibinfo {author} {\bibfnamefont {S.}~\bibnamefont
  {Bose}}, \bibinfo {author} {\bibfnamefont {A.~M.}\ \bibnamefont
  {García-García}}, \bibinfo {author} {\bibfnamefont {M.~M.}\ \bibnamefont
  {Ugeda}}, \bibinfo {author} {\bibfnamefont {J.~D.}\ \bibnamefont {Urbina}},
  \bibinfo {author} {\bibfnamefont {C.~H.}\ \bibnamefont {Michaelis}}, \bibinfo
  {author} {\bibfnamefont {I.}~\bibnamefont {Brihuega}}, \ and\ \bibinfo
  {author} {\bibfnamefont {K.}~\bibnamefont {Kern}},\ }\href@noop {} {\bibfield
   {journal} {\bibinfo  {journal} {Nat. Mater.}\ }\textbf {\bibinfo {volume}
  {9}},\ \bibinfo {pages} {550} (\bibinfo {year} {2010})}\BibitemShut {NoStop}%
\bibitem [{\citenamefont {Mayoh}\ and\ \citenamefont
  {Garc\'{\i}a-Garc\'{\i}a}(2014)}]{Mayoh14-SUC-NanoGranular}%
  \BibitemOpen
  \bibfield  {author} {\bibinfo {author} {\bibfnamefont {J.}~\bibnamefont
  {Mayoh}}\ and\ \bibinfo {author} {\bibfnamefont {A.~M.}\ \bibnamefont
  {Garc\'{\i}a-Garc\'{\i}a}},\ }\href@noop {} {\bibfield  {journal} {\bibinfo
  {journal} {Phys. Rev. B}\ }\textbf {\bibinfo {volume} {90}},\ \bibinfo
  {pages} {134513} (\bibinfo {year} {2014})}\BibitemShut {NoStop}%
\bibitem [{\citenamefont {Greene}\ \emph {et~al.}(1972)\citenamefont {Greene},
  \citenamefont {King}, \citenamefont {Zubeck},\ and\ \citenamefont
  {Hauser}}]{Greene72-Cp-Granular-Al}%
  \BibitemOpen
  \bibfield  {author} {\bibinfo {author} {\bibfnamefont {R.~L.}\ \bibnamefont
  {Greene}}, \bibinfo {author} {\bibfnamefont {C.~N.}\ \bibnamefont {King}},
  \bibinfo {author} {\bibfnamefont {R.~B.}\ \bibnamefont {Zubeck}}, \ and\
  \bibinfo {author} {\bibfnamefont {J.~J.}\ \bibnamefont {Hauser}},\
  }\href@noop {} {\bibfield  {journal} {\bibinfo  {journal} {Phys. Rev. B}\
  }\textbf {\bibinfo {volume} {6}},\ \bibinfo {pages} {3297} (\bibinfo {year}
  {1972})}\BibitemShut {NoStop}%
\bibitem [{\citenamefont {McMillan}(1968)}]{McMillan-Tc-StrongCoupling}%
  \BibitemOpen
  \bibfield  {author} {\bibinfo {author} {\bibfnamefont {W.~L.}\ \bibnamefont
  {McMillan}},\ }\href@noop {} {\bibfield  {journal} {\bibinfo  {journal}
  {Phys. Rev.}\ }\textbf {\bibinfo {volume} {167}},\ \bibinfo {pages} {331}
  (\bibinfo {year} {1968})}\BibitemShut {NoStop}%
\bibitem [{\citenamefont {Allen}\ \emph {et~al.}(1986)\citenamefont {Allen},
  \citenamefont {Beaulac}, \citenamefont {Khan}, \citenamefont {Butler},
  \citenamefont {Pinski},\ and\ \citenamefont
  {Swihart}}]{Allen86-transport-metals}%
  \BibitemOpen
  \bibfield  {author} {\bibinfo {author} {\bibfnamefont {P.~B.}\ \bibnamefont
  {Allen}}, \bibinfo {author} {\bibfnamefont {T.~P.}\ \bibnamefont {Beaulac}},
  \bibinfo {author} {\bibfnamefont {F.~S.}\ \bibnamefont {Khan}}, \bibinfo
  {author} {\bibfnamefont {W.~H.}\ \bibnamefont {Butler}}, \bibinfo {author}
  {\bibfnamefont {F.~J.}\ \bibnamefont {Pinski}}, \ and\ \bibinfo {author}
  {\bibfnamefont {J.~C.}\ \bibnamefont {Swihart}},\ }\href@noop {} {\bibfield
  {journal} {\bibinfo  {journal} {Phys. Rev. B}\ }\textbf {\bibinfo {volume}
  {34}},\ \bibinfo {pages} {4331} (\bibinfo {year} {1986})}\BibitemShut
  {NoStop}%
\bibitem [{\citenamefont {Matthias}\ \emph {et~al.}(1980)\citenamefont
  {Matthias}, \citenamefont {Stewart}, \citenamefont {Giorgi}, \citenamefont
  {Smith}, \citenamefont {Fisk},\ and\ \citenamefont
  {Barz}}]{MATTHIAS80-Enhanced-Tc-YxIr}%
  \BibitemOpen
  \bibfield  {author} {\bibinfo {author} {\bibfnamefont {B.~T.}\ \bibnamefont
  {Matthias}}, \bibinfo {author} {\bibfnamefont {G.~R.}\ \bibnamefont
  {Stewart}}, \bibinfo {author} {\bibfnamefont {A.~L.}\ \bibnamefont {Giorgi}},
  \bibinfo {author} {\bibfnamefont {J.~L.}\ \bibnamefont {Smith}}, \bibinfo
  {author} {\bibfnamefont {Z.}~\bibnamefont {Fisk}}, \ and\ \bibinfo {author}
  {\bibfnamefont {H.}~\bibnamefont {Barz}},\ }\href@noop {} {\bibfield
  {journal} {\bibinfo  {journal} {Science}\ }\textbf {\bibinfo {volume}
  {208}},\ \bibinfo {pages} {401} (\bibinfo {year} {1980})}\BibitemShut
  {NoStop}%
\end{thebibliography}%

\end{document}